\documentclass[a4paper,11pt]{article}
\pdfoutput=1
\usepackage{jcappub}
\usepackage{slashed}

\usepackage{mathtools}
\usepackage[T1]{fontenc}
\usepackage[normalem]{ulem}
\usepackage{pifont}

\newcommand{\gs}{g_\star}
\newcommand{\gss}{g_{\star s}}
\newcommand{\Trh}{T_\text{rh}}
\newcommand{\arh}{a_\text{rh}}
\newcommand{\Tmax}{T_\text{max}}
\newcommand{\amax}{a_\text{max}}
\newcommand{\rGW}{\rho_\text{GW}}
\newcommand{\rR}{\rho_R}
\newcommand{\rp}{\rho_\phi}
\newcommand{\oGW}{\Omega_\text{GW}}
\newcommand{\DNeff}{\Delta N_\text{eff}}

\title{Gravitational Wave from Graviton Bremsstrahlung during Reheating}
\author[a]{Basabendu Barman,}
\author[b]{Nicolás Bernal,}
\author[c]{\\Yong Xu,}
\author[d]{and Óscar Zapata}

\affiliation[a]{Institute of Theoretical Physics, Faculty of Physics, University of Warsaw\\
ul. Pasteura 5, 02-093 Warsaw, Poland}
\affiliation[b]{New York University Abu Dhabi\\
PO Box 129188, Saadiyat Island, Abu Dhabi, United Arab Emirates}
\affiliation[c]{PRISMA+ Cluster of Excellence and Mainz Institute for Theoretical Physics\\
Johannes Gutenberg University, 55099 Mainz, Germany}
\affiliation[d]{Instituto de Física, Universidad de Antioquia\\
Calle 70 \# 52-21, Apartado Aéreo 1226, Medellín, Colombia}

\emailAdd{basabendu88barman@gmail.com}
\emailAdd{nicolas.bernal@nyu.edu}
\emailAdd{yonxu@uni-mainz.de}
\emailAdd{oalberto.zapata@udea.edu.co}

\abstract{We revisit graviton production via Bremsstrahlung from the decay of the inflaton during inflationary reheating. Using two complementary computational techniques, we first show that such 3-body differential decay rates differ from previously reported results in the literature. We then compute the stochastic gravitational wave (GW) background that forms during the period of reheating, when the inflaton perturbatively decays with the radiative emission of gravitons. By computing the number of relativistic degrees of freedom in terms of $\DNeff$, we constrain the resulting GW energy density from BBN and CMB. Finally, we project current and future GW detector sensitivities in probing such a stochastic GW background, which typically peaks in the GHz to THz ballpark, opening up the opportunity to be detected with microwave cavities and space-based GW detectors.}

\begin{document}
\begin{flushright}
\end{flushright}

\maketitle

\section{Introduction}
The existence of a primordial gravitational wave (GW) background is one of the most crucial predictions of the inflationary scenario of the early universe. Stochastic GWs can have several sources, {\it viz.,} from the quantum fluctuations during inflation~\cite{Starobinsky:1979ty, Allen:1987bk, Sahni:1990tx, Turner:1993vb} that give rise to tensor perturbations, during preheating~\cite{Khlebnikov:1997di, Easther:2006gt, Dufaux:2007pt, Bethke:2013aba, Figueroa:2017vfa} when rapid particle production via parametric resonance occurs or from oscillations of cosmic string loops~\cite{Vilenkin:1981bx, Cui:2017ufi, Cui:2018rwi, Chang:2021afa}, originated from, for example, a spontaneously broken $U(1)$ symmetry (gauged or global) or from the standard model (SM) plasma in thermal equilibrium~\cite{Ghiglieri:2015nfa, Ghiglieri:2020mhm, Ghiglieri:2022rfp}.
However, as pointed out in Refs.~\cite{Nakayama:2018ptw, Huang:2019lgd}, stochastic GWs of primordial origin can be sourced from the decay of the inflaton.\footnote{Such graviton can also act as a mediator in the production of the dark matter relic abundance~\cite{Mambrini:2022uol}.}
In that case, after the end of inflation, during the era of reheating, the inflaton field can decay into particles of arbitrary spins, depending on the microscopic nature of its interaction. Considering gravitons to emerge as quantum fluctuations over the classical background, they inexorably couple to matter, leading to a graviton production from inflaton decays, similar to the Bremsstrahlung process as considered in Ref.~\cite{Weinberg:1965nx}. It is then unavoidable to have inflaton decay as a source of the primordial GW background. 

With this motivation, in this work, we revisit the scenario in which the inflaton can interact with bosons or fermions, leading to its {\it perturbative} decay during reheating, resulting in the production of a SM radiation bath. Here, we would like to emphasize that inflaton decay via trilinear couplings fully drains the inflaton energy, allowing the Universe to transit into a radiation-domination phase~\cite{Dufaux:2006ee}. By considering fluctuations over a flat background, we introduce the dynamical (massless) graviton field of spin 2 that communicates with all other matter fields through the energy-momentum tensor. This eventually leads to 3-body decay of the inflaton, involving a pair of scalars, fermions, or vector bosons, along with the radiative emission of a graviton. In computing the 3-body decay widths, we follow two complementary approaches: $a)$ we explicitly construct the graviton polarization tensors, and $b)$ we utilize the polarization sum and show that our expressions converge in either case, however, differing from previous analyses reported in Refs.~\cite{Nakayama:2018ptw, Huang:2019lgd, Ghoshal:2022kqp}. It is then possible to compute the GW energy density from the differential 3-body decay width of the inflaton. 

As is well known, in order for Big Bang Nucleosynthesis (BBN) to proceed successfully, the energy budget of the Universe must not comprise a significant amount of extra relativistic species, including GWs. This condition requires that the energy fraction of GWs to the SM radiation degrees of freedom (DoF) at that time is not greater than about $\sim 10\%$. Regardless of its origin, the energy density in GW established before BBN acts as radiation, and thus its impact on BBN is fully captured by $\DNeff$, which counts the number of relativistic species. Furthermore, GWs with initial adiabatic conditions leave the same imprint on the CMB as free-streaming dark radiation, and in this case, the limit on the present-day energy density in GWs is $\oGW^{(0)}\, h^2 < 1.3 \times 10^{-6}$~\cite{Pagano:2015hma, Caprini:2018mtu}. We discuss the impact of the CMB measurement of $\DNeff$ on the GW energy density emitted from the decay of the inflaton, taking into account the evolution of the energy densities during reheating. We compare the predicted spectrum of stochastic GWs with existing and future experiments, finding that the present GW spectrum strongly requires high-frequency GW detectors. Interestingly, we see that such high-frequency GWs could be detected, for example, with resonant cavity detectors~\cite{Berlin:2021txa, Herman:2022fau} or with space-based futuristic GW detectors~\cite{Seto:2001qf, Kudoh:2005as}. We refer to Ref.~\cite{Aggarwal:2020olq} for a recent review on  high-frequency GW searches.

The paper is organized as follows. We present the underlying interaction Lagrangian and present the 2- and 3-body decay rates in Section~\ref{sec:franework}. In Section~\ref{sec:DNeff} we calculate the constraints from $\DNeff$ on the GW energy density. The computation of the primordial GW spectrum is presented in Section~\ref{sec:GWspectrum}. Finally, we conclude in Section~\ref{sec:concl}. In the appendixes, we present our calculations in detail.

\section{The Framework} \label{sec:franework}
The underlying interaction Lagrangian for the present set-up can be divided into two parts. One part involving a trilinear interaction between the inflaton $\phi$ and a pair of complex scalar doublets $\varphi$ with 4 DoF (which is the SM Higgs field), a pair of vector-like Dirac fermions $\psi$ with 4 DoF, or a pair of massive vector bosons $V_\mu$ with 3 DoF, given by
\begin{equation}\label{eq:int1}
    \mathcal{L}_\text{int}^{(2)} \supset -\mu\, \phi\, |\varphi|^2 - y_\psi\, \overline{\psi}\, \psi\, \phi - g_V\, V_\mu\, V^\mu\, \phi \,,    
\end{equation}
where the corresponding interaction strengths are parameterized in terms of the couplings $\mu$, $y_\psi$, and $g_V$, respectively. The superscript $(2)$ denotes interactions that lead to a two-body decay of the inflaton. Also, note that the coupling strength $\mu$ and $g_V$ have mass dimension, while the Yukawa interaction strength $y_\psi$ is dimensionless. Here, we remain agnostic about the underlying UV-complete Lagrangian and, for simplicity, work with an effective theory. 

On the other hand, since we are interested in the {\it unavoidable} Bremsstrahlung process involving gravitons, we expand the metric $g_{\mu \nu}$ around Minkowski spacetime: $g_{\mu \nu} \simeq \eta_{\mu \nu} + \frac{2}{M_P}\, h_{\mu \nu}$, where $M_P$ is the reduced Planck mass.
This inevitably leads to gravitational interactions that are described by the Lagrangian~\cite{Choi:1994ax}
\begin{equation} \label{eq:int2}
    \sqrt{-g}\, \mathcal{L}_{\rm int}^{(g)} \supset -\frac{2}{M_P}\, h_{\mu \nu}\, T^{\mu \nu},
\end{equation}
where $h_{\mu\nu}$ refers to the graviton field that appears as a quantum fluctuation on the flat background, and $T_{\mu\nu}$ represents the energy-momentum tensor involving {\it all} matter particles involved in the theory. Further, we do not consider any non-minimal coupling between the new fields of the theory and gravity; hence, this is a minimal scenario. All relevant Feynman rules involving the graviton and particles of different spins (0, 1/2, and 1) are elaborated in Appendix~\ref{sec:app-vertex}. The interactions appearing in Eqs.~\eqref{eq:int1} and~\eqref{eq:int2} give rise to 2- and 3-body decays of the inflaton into pairs of $\varphi$, $\psi$, and $V$ in the final state, along with the emission of a massless graviton. After production, gravitons propagate and constitute the stochastic GW background, the spectrum of which we shall compute, considering different spins of the final-state products.

With this setup, we now move on to the discussion of three different decay scenarios, where the inflaton $\phi$ perturbatively decays into either a pair of bosons or a pair of fermions, with graviton radiation, due to the graviton-matter coupling. In the following sections, we discuss three cases individually.

Before proceeding, we would like to comment on the possible non-perturbative preheating~\cite{Kofman:1997yn, Drewes:2019rxn, Drewes:2022nhu}. For the bosonic case, due to the trilinear coupling, the daughter particles, i.e. the Higgs bosons, could feature a tachyonic mass $\sim \mu\, \phi$, leading to non-perturbative particle production~\cite{Dufaux:2006ee}. However, in our case, the bosonic decay product is the SM Higgs, which has a sizable self-coupling $\lambda\, (\varphi^\dagger \varphi)^2$ that induces a positive mass term $\sim \lambda  \left\langle  \varphi^2 \right  \rangle$ (with $\left\langle  \varphi^2 \right  \rangle$ being the Higgs variance) once the Higgses are copiously produced. This backreaction counteracts the tachyonic mass and quickly terminates non-perturbative particle production, making preheating less efficient~\cite{Drees:2021wgd}. On the other hand, for the fermionic channel, the Pauli blocking effect implies that only a small fraction of the energy stored in the inflaton field can be transferred non–perturbatively~\cite{Peloso:2000hy}. Thereafter, in our setup, a perturbative treatment could capture the dominant phenomena occurring in the reheating phase. We also mention that in the literature, scenarios such as instant preheating~\cite{Felder:1998vq} could deplete the inflaton energy more efficiently. 

\subsection{Decay into Scalars} \label{sec:scalar}
We start with the inflaton decay into spin-0 states, where the final-state particles are considered to be complex doublet scalars, e.g. the SM Higgs doublet. The 2-body decay rate in this case, following the Lagrangian in Eq.~\eqref{eq:int1}, is given by
\begin{equation}\label{eq:Gamma0scalar}
    \Gamma_0^{(0)} = \frac{2\, M}{16\,\pi} \left(\frac{\mu}{M}\right)^2 \sqrt{1 - 4\, y^2}\,,
\end{equation}
where $y \equiv m/M$, with $m$ being the mass of the daughter particles (independent of their spin). The factor of 2 appears due to two possible decay channels for the complex scalar doublet in the final state. The subscript represents the spin of the final-state particles, while the superscript $(0)$ denotes the 2-body decay width. 
\begin{figure}[t!]
    \centering
    \includegraphics[scale=0.14]{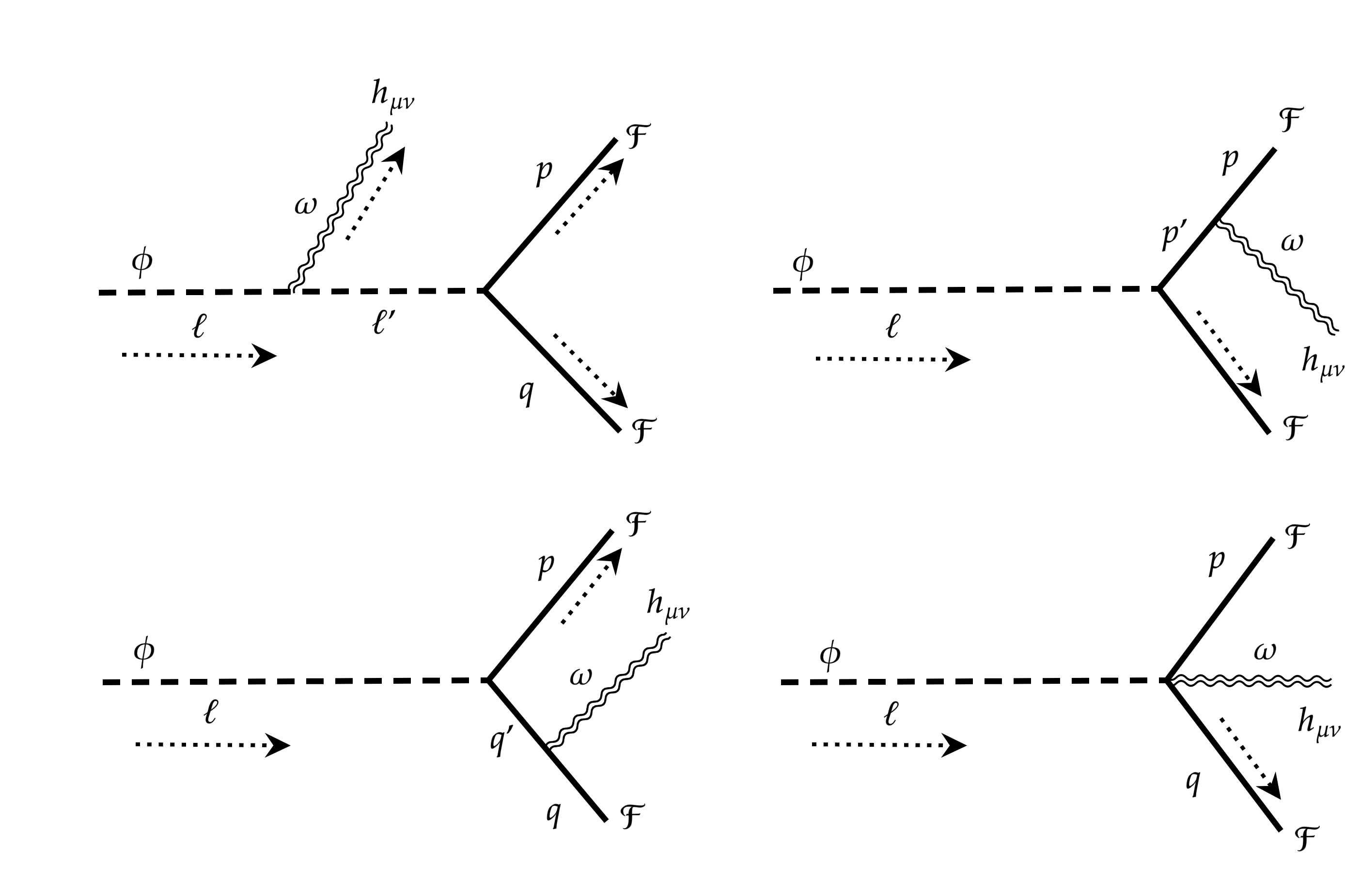}
    \caption{Feynman diagrams for an inflaton decay into a pair of particles $\mathcal{F}$, along with a radiated graviton. Here $\mathcal{F}$ could be a scalar $\varphi$, a fermion $\psi$, or a vector $V$, while $h_{\mu\nu}$ is the graviton tensor field. We denote the incoming and outgoing momenta with dashed arrowheads.}
    \label{fig:feyn}
\end{figure} 

As advocated before, due to the irreducible gravitational interaction (cf. Eq.~\eqref{eq:int2}), the final state could also contain a graviton~\cite{Nakayama:2018ptw}, leading to a 3-body decay of $\phi$. The general three-body decay diagrams are shown in Fig.~\ref{fig:feyn}, with $l$, $\omega$, $p$, and $q$ denoting the initial and final four-momentum, respectively.\footnote{The amplitude of the bottom right diagram is proportional to $\eta_{\mu \nu} \epsilon^{\mu \nu}$ (cf. Eq.~\eqref{eq:traceless}) and therefore vanishes due to the traceless condition for a massless graviton.} Here, we denote any general final state as $\mathcal{F}$, where $\mathcal{F}$ can be a scalar, a fermion, or a gauge boson. The detailed computation of the 3-body decay following two different methodologies, namely the explicit construction of graviton polarization tensors and the polarization sum, is reported in Appendix~\ref{sec:app-decay}. We would like to mention here that, in defining the polarization of massless gravitons, we have taken care of the fact that they satisfy the transverse, traceless, symmetric, and orthonormal conditions in either of the methodologies mentioned above. This results in a different final expression for the 3-body decay width from those derived in Ref.~\cite{Nakayama:2018ptw}. The differential decay rate for the scalar final state with the emission of a graviton of energy $E_\omega$ reads
\begin{equation} \label{eq:dGdE}
    \frac{d\Gamma_0^{(1)}}{dE_\omega} = \frac{1}{32\,\pi^3} \left(\frac{\mu}{M_P}\right)^2 \left[\frac{(1 - 2 x)\, (1 - 2 x + 2 y^2)}{4 x\, \alpha^{-1}} + \frac{y^2\, (y^2 + 2 x - 1)}{x} \ln\left(\frac{1 + \alpha}{1 - \alpha}\right)\right], 
\end{equation}
with $x \equiv E_\omega/M$ and 
\begin{equation}
    \alpha \equiv \sqrt{1 - \frac{4\, y^2}{1-2x}}\,,
\end{equation}
with a graviton energy spanning the range
\begin{equation}
    0 < E_\omega \leq M \left(\frac12 - 2\, y^2\right).
\end{equation}
Since the differential rate in Eq.~\eqref{eq:dGdE} plays a key role in our subsequent calculation, we would like to make some remarks before proceeding. Note that a graviton could carry at most half of the inflaton energy, which occurs in a limit where the daughter particle mass is zero, namely $y \to 0$. In such a case, the differential decay rate vanishes as the phase space closes. More generally, the differential decay rate should also vanish when $x \to \frac12 - 2\, y^2$. We notice that our result differs from that reported in Eq.~(7) of Ref.~\cite{Nakayama:2018ptw}.

\subsection{Decay into Fermions} \label{sec:fermion}
Following the second term in Eq.~\eqref{eq:int1}, we compute the 2-body decay of $\phi$ into a pair of fermions in the final state. In that case, the decay width is given by
\begin{equation}
    \Gamma_{1/2}^{(0)} = \frac{y_\psi^2}{8\pi}\, M \left(1-4\,y^2\right)^{3/2}.
\end{equation}
As before, one can compute the differential rate of the three-body final state involving a pair of $\psi$'s and a graviton in the final state, leading to
\begin{align}
    \frac{d\Gamma_{1/2}^{(1)}}{dE_\omega} = \frac{y_\psi^2}{64\,\pi^3} \left(\frac{M}{M_P}\right)^2\Bigg[&\frac{1 - 2 x}{x\, \alpha^{-1}} \left[8 x\, y^2 + 2 x\, (x - 1) - 8 y^4 - 2 y^2 + 1\right] \nonumber\\
    &+ \frac{4\,y^2 \left[(5 - 8 x)\, y^2 - (x - 1)^2 - 4 y^4\right]}{x} \ln\left(\frac{1 + \alpha}{1 - \alpha}\right) \Bigg],
\end{align}
see Appendix~\ref{sec:app-ferm} for details.
Interestingly, we again find that our expression for the 3-body decay rate differs from those reported in Eq.~(8) of Ref.~\cite{Nakayama:2018ptw} and Eq.~(B.1) of Ref.~\cite{Ghoshal:2022kqp}. 

\subsection{Decay into Vectors}
For inflaton decays to massive vectors\footnote{We notice that a similar process was also considered in Ref.~\cite{Klose:2022knn}, where the inflaton was assumed to be an axion-like particle.} via the trilinear interaction term $\phi\, V_\mu V^{\mu}$, the 2-body decay rate is given by 
\begin{equation} \label{eq:GV}
    \Gamma_1^{(0)} = \frac{M}{64\,\pi} \left(\frac{g_V}{M}\right)^2\, \frac{1-4\,y^2+12\,y^4}{y^4}\, \sqrt{1-4\,y^2}\,,
\end{equation}
while the 3-body differential decay rate reads (see Appendix~\ref{sec:app-vector} for details of the computation)
\begin{align} \label{eq:dGVdE}
    \frac{d\Gamma^{(1)}_1}{dE_\omega} &= \frac{1}{1920\,\pi^3\,x\,y^4} \left(\frac{g_V}{M_P}\right)^2 \bigg(\alpha\,\Big[360\,(1-2 x)\,y^6+4\,\left(4 x\,\left(23\,x-5\right)+15\right)\,y^4 \nonumber\\
    &+ 2\,(2\,x-1)\,(28\,x\,(14\,x-5)+15)\,y^2 +(1-2 x)^2\,\left(4\,x\,\left(2\,x-5\right)+15\right) \Big] \nonumber\\
    &+ 60\,y^2 \left[12 y^6 + 16 (x-1) y^4 + (5 + 4x (4x-3)) y^2 - ( 1-2x)^2 (1+2x)\right] \ln\left[\frac{1 + \alpha}{1 - \alpha}\right]\bigg)\,.
\end{align}
Note that factor $1/y^4$ comes from the polarization sum for the massive vector and therefore the massless case cannot be recovered in the limit $y \to 0$.
We would like to mention that our results in Eqs.~\eqref{eq:GV} and~\eqref{eq:dGVdE} differ from the ones reported in Eqs.~(4) and~(7) of Ref.~\cite{Huang:2019lgd}.

\section{\boldmath Gravitational Wave Contribution to $\DNeff$} \label{sec:DNeff}
As we know, to switch to the standard hot Big Bang cosmology after inflation, the inflaton energy must be transferred into SM radiation DoFs, which eventually thermalize and dominate the Universe's energy budget. This transition process, known as {\it reheating}, is typically marked by the equality between the inflaton and radiation energy densities. The reheating process must end before the onset of BBN, which occurs at $T_\text{BBN} \simeq 4$~MeV~\cite{Sarkar:1995dd, Kawasaki:2000en, Hannestad:2004px, DeBernardis:2008zz, deSalas:2015glj}. Now, in order for BBN to proceed successfully, the energy budget of the Universe must not comprise a significant amount of extra relativistic species, including GWs. Regardless of its origin, the energy density established in GW before BBN acts as radiation, and thus its impact on BBN is fully captured in terms of $\DNeff$. Therefore, an excess of the GW energy density around BBN can be restricted by considering the (present and future) bounds on $\DNeff$ from CMB, BBN, and combined. In this section, we discuss our calculations considering that the inflaton $\phi$ oscillates in a simple quadratic potential, which implies that its energy density scales as non-relativistic matter during reheating.

The number of effective neutrinos $N_\text{eff}$ is defined from the expression of the radiation energy density in the late universe (at a photon temperature $T_{\DNeff}$) as
\begin{equation}
    \rho_\text{rad}(T_{\DNeff}) = \rho_\gamma + \rho_\nu + \rGW = \left[1 + \frac78 \left(\frac{T_\nu}{T_\gamma}\right)^4 N_\text{eff}\right] \rho_\gamma(T_{\DNeff})\,,
\end{equation}
where $\rho_\gamma$, $\rho_\nu$, and $\rGW$ correspond to the photon, SM neutrino, and GW energy densities, respectively, with $T_\nu/T_\gamma = (4/11)^{1/3}$. Note that $T_{\DNeff}$ corresponds to the temperature at which the effective number of neutrinos is evaluated. There are experimental bounds on $\DNeff$ for $T_{\DNeff} = T_\text{BBN}$ and $T_{\DNeff} = T_\text{CMB}$, where $T_\text{CMB}$ is the temperature at which the photons decouple from the thermal plasma. Within the SM, the prediction taking into account the non-instantaneous neutrino decoupling is $N_\text{eff}^\text{SM} = 3.044$~\cite{Dodelson:1992km, Hannestad:1995rs, Dolgov:1997mb, Mangano:2005cc, deSalas:2016ztq, EscuderoAbenza:2020cmq, Akita:2020szl, Froustey:2020mcq, Bennett:2020zkv}, whereas the presence of GWs implies
\begin{equation} \label{eq:DNeff}
    \DNeff \equiv N_\text{eff} - N_\text{eff}^\text{SM} = \frac87 \left(\frac{11}{4}\right)^\frac43 \frac{\rGW(T_{\DNeff})}{\rho_\gamma(T_{\DNeff})} = \frac87 \left(\frac{11}{4} \frac{\gss(T_{\DNeff})}{\gss(\Trh)}\right)^\frac43 \frac{\gs(\Trh)}{2} \frac{\rGW(\Trh)}{\rR(\Trh)},
\end{equation}
where
\begin{align}
    \rR(T) &= \frac{\pi^2}{30}\, \gs(T)\, T^4,\\
    s(T) &= \frac{2 \pi^2}{45}\, \gss(T)\, T^3
\end{align}
are the SM radiation energy density and the SM entropy density, with $\gs(T)$ and $\gss(T)$ the numbers of relativistic degrees of freedom~\cite{Drees:2015exa}.

The evolution of inflaton, SM radiation, and GW energy densities can be tracked using the Boltzmann equations\footnote{We would like to emphasize that our approach of computation of the GW energy density takes care of the evolution of energy densities beyond the instantaneous approximation as was done in Refs.~\cite{Nakayama:2018ptw, Huang:2019lgd}.}\footnote{We note that the term proportional to $\Gamma^{(1)}$ in the RHS of Eq.~\eqref{eq:beq1} can be rewritten as
\begin{equation*}
    \Gamma^{(1)}\rho_\phi = \int \frac{d\Gamma^{(1)}}{dE_\omega}  dE_\omega \,\rho_\phi = \int \frac{d\Gamma^{(1)}}{dE_\omega}\, \frac{M - E_\omega}{M}\, dE_\omega \,\rho_\phi  + \int   \frac{d\Gamma^{(1)}}{dE_\omega}\, \frac{ E_\omega}{M}\, dE_\omega \,\rho_\phi\,.
\end{equation*}
This expression, even if less compact, allows to easily check that the sum of the RHS of Eqs.~\eqref{eq:beq1}, \eqref{eq:beq2}, and~\eqref{eq:beq3} is zero, implying conservation of the energy density.}
\begin{align}
    &\frac{d\rp}{dt} + 3\, H\, \rp = - \left(\Gamma^{(0)} + \Gamma^{(1)}\right) \rp\,,\label{eq:beq1}\\
    &\frac{d\rR}{dt} + 4\, H\, \rR = + \Gamma^{(0)}\, \rp + \int \frac{d\Gamma^{(1)}}{dE_\omega}\, \frac{M - E_\omega}{M}\, \rp\, dE_\omega\,, \label{eq:beq2}\\
    &\frac{d\rGW}{dt} + 4\, H\, \rGW = + \int \frac{d\Gamma^{(1)}}{dE_\omega}\, \frac{E_\omega}{M}\, \rp\, dE_\omega\,, \label{eq:beq3}
\end{align}
where $H$ stands for the Hubble expansion rate given by
\begin{equation}
    H^2 = \frac{\rp + \rR + \rGW}{3\, M_P^2}\,,
\end{equation}
while $\Gamma^{(0)}$ and $\Gamma^{(1)}$ are the 2- and 3-body decay widths.
The factors $(M - E_\omega)/M$ and $E_\omega/M$ correspond to the fractions of inflaton energy injected into SM radiation and GWs, respectively.
It follows that
\begin{equation} \label{eq:BEratio}
    \frac{d (\rGW / \rR)}{da} \simeq \frac{1}{a\, H}\, \frac{\rp}{\rR} \left[\int \frac{d\Gamma^{(1)}}{dE_\omega}\, \frac{E_\omega}{M}\, dE_\omega - \frac{\rGW}{\rR}\, \Gamma^{(0)}\right].
\end{equation}
This expression can be integrated during reheating, that is, for $\amax \leq a \leq \arh$, corresponding to photon temperatures $\Tmax \geq T \geq \Trh$. Note that, in solving Eq.~\eqref{eq:BEratio}, we have considered $\rGW/\rR$ as a variable itself, which makes Eq.~\eqref{eq:BEratio} an ordinary first-order differential equation. Importantly, during reheating in which the SM thermal bath is produced and the universe transitions to radiation domination, the bath temperature may rise to a value $\Tmax$ that exceeds $\Trh$~\cite{Giudice:2000ex}. The possibility that the maximum temperature of the thermal bath may reach $\Tmax > \Trh$ before cooling is not apparent if one takes the instantaneous decay approximation for reheating. We note that during reheating
\begin{align}
    \rp(a) &= \rp(\arh) \left(\frac{\arh}{a}\right)^3,\label{eq:rp}\\
    T(a) &= \Trh \left(\frac{\arh}{a}\right)^{3/8},\label{eq:T}
\end{align}
as the inflaton is assumed to be non-relativistic and to decay with a constant decay width into SM radiation. The solutions in Eqs.~\eqref{eq:rp} and~\eqref{eq:T} can be realized from Eqs.~\eqref{eq:beq1} and~\eqref{eq:beq2}, considering the fact that during the early stage of reheating (inflaton domination), the decay rate of the inflaton is much smaller than the expansion rate~\cite{Giudice:2000ex}, and also utilizing the fact that $\Gamma^{(0)} \gg \Gamma^{(1)}$. We emphasize that the scaling of the SM temperature is due to the fact that the SM radiation is not free, but is sourced by inflaton decays. The end of the reheating corresponds to the moment in which the equality $\rR(\Trh) = \rp(\Trh)$ is realized. Additionally, assuming that at the beginning of the reheating, the universe had no SM radiation or GWs, and taking into account that at the end of the reheating $\Gamma^{(0)} \simeq H(\Trh)$, Eq.~\eqref{eq:BEratio} admits the analytical solution
\begin{equation} \label{eq:ratiorho}
    \frac{\rGW(\Trh)}{\rR(\Trh)} \simeq \int_0^{M/2} \frac{1}{\Gamma^{(0)}}\, \frac{d\Gamma^{(1)}}{dE_\omega}\, \frac{E_\omega}{M}\, dE_\omega \left[1 - \left(\frac{\Trh}{\Tmax}\right)^{8/3}\right].
\end{equation}
We notice that within the approximation of an instantaneous decay of the inflaton, the expression in the squared brackets reduces to one.

For the different decay channels, in the limit $y\to 0$, one has
\begin{equation} \label{eq:ratioDNeff}
     \frac{\rGW(\Trh)}{\rR(\Trh)} \simeq \mathcal{C}_\rho\, \frac{M^2}{\pi^2 M_P^2} \left[1 - \left(\frac{\Trh}{\Tmax}\right)^{8/3}\right],
\end{equation}
where $\mathcal{C}_\rho = 1/96$ for scalars, $3/128$ for fermions, and $127/1800$ for vectors.
Therefore, the corresponding GW contribution to $\DNeff$ is
\begin{equation}\label{eq:DNeff2}
    \DNeff \simeq \mathcal{C}_{\DNeff} \left(\frac{M}{M_P}\right)^2 \left[1 - \left(\frac{\Trh}{\Tmax}\right)^{8/3}\right],
\end{equation}
with $\mathcal{C}_{\DNeff} \simeq 0.01$ for scalars, 0.03 for fermions, and 0.08 for vectors, where we have taken $\gss(T_{\DNeff}) \simeq 10.75$, considering $T_{\DNeff} = T_\text{BBN}$. Again, note that in the instantaneous reheating approximation, the square bracket in Eq.~\eqref{eq:DNeff2} becomes unity.  To avoid jeopardizing the successful predictions from BBN, the reheating temperature must satisfy $\Trh \geq T_\text{BBN}$. Furthermore, recent BICEP/Keck measurements have offered a stronger bound (than that of previous Planck results~\cite{Planck:2018vyg}) on the tensor-to-scalar ratio $r < 0.035$~\cite{BICEP:2021xfz}, implying $\Trh \lesssim 5.5 \times 10^{15}$~GeV.

\begin{figure}[t!]
    \centering
    \includegraphics[scale=0.85]{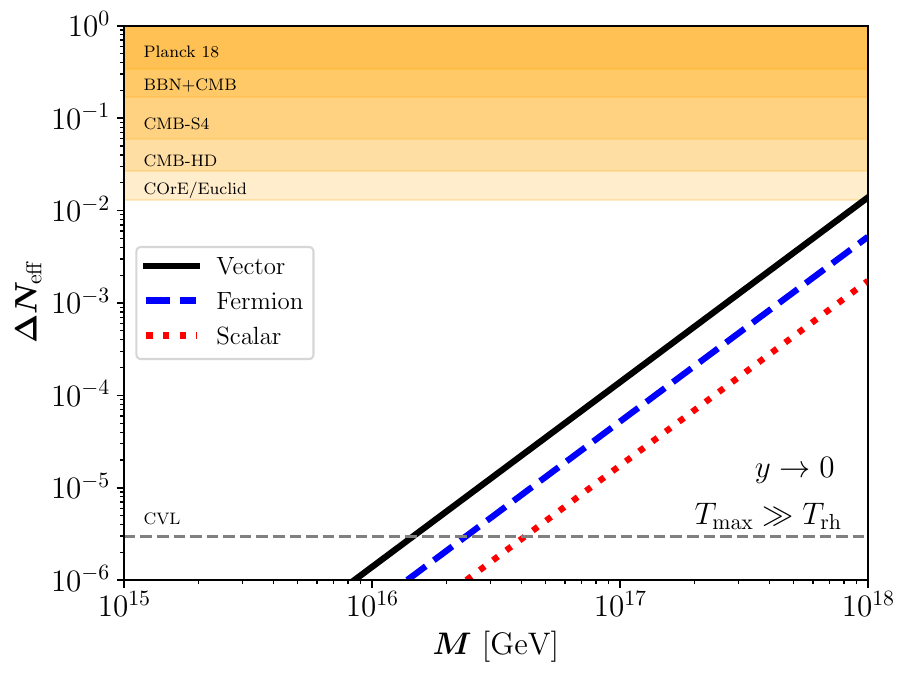}
    \caption{Contribution of GW energy density to $\DNeff$ (cf. Eq.~\eqref{eq:DNeff2} with $\Tmax \gg \Trh$), where 
    the solid black, dashed blue, and dot-dashed red slanted straight lines correspond to scalar, fermion, and vector boson final states. We show the present limits from 
    PLANCK~\cite{Planck:2018vyg}, CMB+BBN combined~\cite{Yeh:2022heq}, and future limits from CMB-S4~\cite{Abazajian:2019eic},  CMB-HD~\cite{CMB-HD:2022bsz}, COrE~\cite{COrE:2011bfs}/ Euclid~\cite{EUCLID:2011zbd}, and also hypothetical CVL experiment~\cite{Ben-Dayan:2019gll}, from top to bottom.}
    \label{fig:DNeff}
\end{figure} 
Within the framework of $\Lambda$CDM, Planck legacy data produces $N_\text{eff} = 2.99 \pm 0.34$ at 95\% CL~\cite{Planck:2018vyg}. Once the baryon acoustic oscillation (BAO) data are included, the measurement becomes more stringent: $N_\text{eff} = 2.99 \pm 0.17$ at 1$\sigma$ CL. The upcoming CMB experiments, such as SPT-3G~\cite{SPT-3G:2014dbx} and the Simons Observatory~\cite{SimonsObservatory:2018koc}, will soon improve Planck's precision on $N_\text{eff}$. In particular, CMB-S4~\cite{Abazajian:2019eic} and CMB-HD~\cite{CMB-HD:2022bsz} will be sensitive to a precision of $\DNeff \sim 0.06$ and $\DNeff \sim 0.027$ at 95\% CL, respectively. As calculated in Ref.~\cite{Yeh:2022heq}, a combined analysis from BBN and CMB results in $N_\text{eff} = 2.880 \pm 0.144$. The next generation of satellite missions, such as COrE~\cite{COrE:2011bfs} and Euclid~\cite{EUCLID:2011zbd}, shall impose limits at $2\sigma$ on $\DNeff \lesssim 0.013$. Furthermore, as mentioned in Ref.~\cite{Ben-Dayan:2019gll}, a hypothetical cosmic-variance-limited (CVL) CMB polarization experiment could presumably be reduced to as low as $\DNeff \lesssim 3 \times 10^{-6}$, although this does not seem to be an experimentally plausible scenario. Fig.~\ref{fig:DNeff} illustrates the constraint from $\DNeff$ following Eq.~\eqref{eq:DNeff2}, considering $\Tmax \gg \Trh$. As discussed above, we show the present and future limits of $\DNeff$ on the GW energy density for scenarios in which the graviton decays into a pair of scalars (red dotted line), a pair of Dirac fermions (blue dashed line), or a pair of massive vector bosons (black solid line). As we can see, the impact of GW production on $\DNeff$ through all these channels is very challenging not only for present, but even for the projected experimental sensitivities, unless $M \sim M_P$. A large inflaton mass is required to overcome the strong Planck suppression originating from minimal graviton coupling. Note that there is a possibility for experiments such as COrE or Euclid to probe the vector scenario. 

\section{Gravitational Wave Spectrum} \label{sec:GWspectrum}
After being produced from inflaton 3-body decays, gravitons would propagate and spread in the whole universe, forming a homogeneous and isotropic stochastic GW background at present, after the attenuation of its energy and amplitude due to cosmic expansion. The primordial GW spectrum at present $\oGW(f)$ for a frequency $f$ is defined by
\begin{equation} \label{eq:oGW}
    \oGW(f) = \frac{1}{\rho_c}\, \frac{d\rGW}{d\ln f} = \Omega_\gamma^{(0)}\, \frac{d(\rGW/\rR)}{d\ln f} = \Omega_\gamma^{(0)}\, \frac{\gs(\Trh)}{\gs(T_0)} \left[\frac{\gss(T_0)}{\gss(\Trh)}\right]^{4/3}\,\frac{d(\rGW(\Trh)/\rR(\Trh))}{d\ln E_\omega}\,,
\end{equation}
where $\rho_c$ is the critical energy density, and $\Omega_\gamma^{(0)} h^2 \simeq 2.47 \times 10^{-5}$ is the observed photon abundance~\cite{Planck:2018vyg}.
Equation~\eqref{eq:oGW} must be evaluated at an energy 
\begin{equation}\label{eq:frequency}
    E_\omega = 2\pi\, f\, \frac{a_0}{\arh} = 2\pi\, f\, \frac{\Trh}{T_0} \left[\frac{\gss(\Trh)}{\gss(T_0)}\right]^{1/3},
\end{equation}
taking into account the redshift of the GW energy between the end of reheating and the present epoch.

From Eq.~\eqref{eq:BEratio}, the evolution of the differential ratio of GW to SM radiation energy densities is given by
\begin{equation}
    \frac{d}{da} \frac{d (\rGW / \rR)}{dE_\omega} \simeq \frac{1}{a\, H(a)}\, \frac{\rp(a)}{\rR(a)} \left[\left.\frac{d\Gamma^{(1)}}{dE'_\omega}\, \frac{E'_\omega}{M}\right|_{E'_\omega = E_\omega \frac{\arh}{a}} - \frac{d (\rGW / \rR)}{dE_\omega}\, \Gamma^{(0)}\right],
\end{equation}
where in the source term for GWs the redshift of the graviton energy $E'_\omega(a) = E_\omega\, \frac{\arh}{a}$ was taken into account. An approximate solution is
\begin{equation}
    \frac{d(\rGW(\Trh)/\rR(\Trh))}{d E_\omega} \simeq \frac{d\Gamma^{(1)}}{dE_\omega}\, \frac{E_\omega}{M}\, \frac{1}{\Gamma^{(0)}} \left[1 -  \left(\frac{\Trh}{\Tmax}\right)^{8/3}\right],
\end{equation}
where again, within the approximation of an instantaneous decay of the inflaton, the expression in the squared brackets reduces to one.
For the inflaton decay into particles with different spins, one has
\begin{equation} \label{eq:oGW2}
    \oGW(f) \simeq \mathcal{C}_{\oGW}\, \frac{\Trh}{5.5\times 10^{15}~\text{GeV}}\, \frac{M}{M_P}\, \frac{f}{10^{12}~\text{Hz}}\,,
\end{equation}
with $\mathcal{C}_{\oGW} \simeq 1.4\times 10^{-8}$ for scalars, $\mathcal{C}_{\oGW} \simeq 2.8 \times 10^{-8}$ for fermions, and $\mathcal{C}_{\oGW} \simeq 11.2 \times 10^{-8}$ for vectors.
The latter expression is valid for frequencies $f$ smaller than
\begin{equation} \label{eq:f}
    f \lesssim  \frac{M}{4\pi}\, \frac{T_0}{\Trh} \left[\frac{\gss(T_0)}{\gss(\Trh)}\right]^{1/3} \simeq 4.1 \times 10^{12}\,\left( \frac{M}{M_P}\right)\, \left(\frac{5.5\times 10^{15}~\text{GeV}}{\Trh}\right) \text{Hz}\,,
\end{equation}
where we have used $\gss(T_0) = 3.94$ and $\gss(\Trh) = 106.75$.

\begin{figure}[t!]
    \centering
    \includegraphics[scale=0.99]{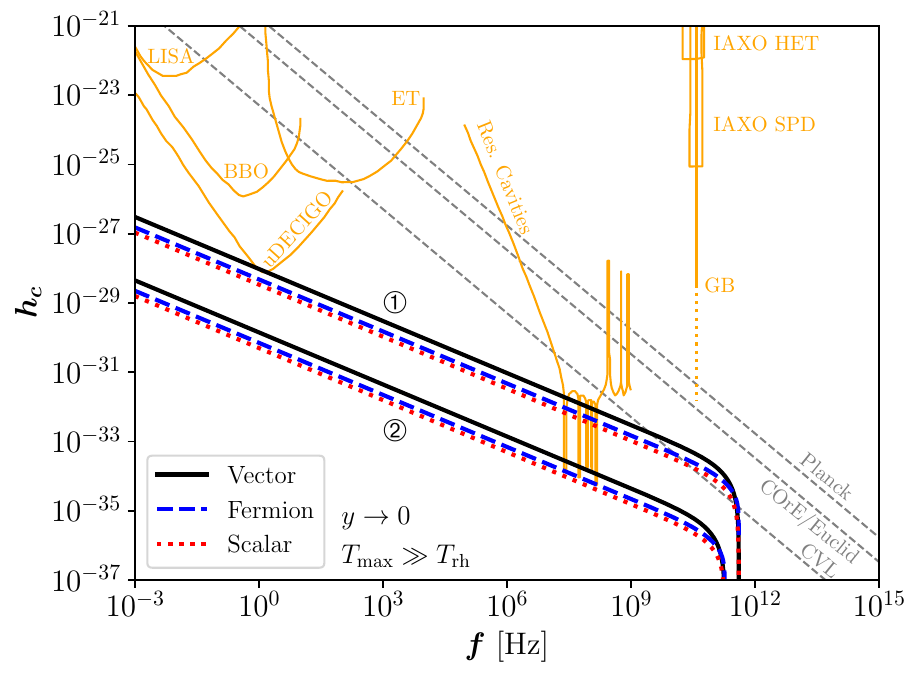}
    \caption{Dimensionless strain $h_c$ as a function of the GW frequency $f$, for the two benchmarks (\ding{192} and \ding{193}) described in the text, assuming $\Tmax \gg \Trh$ and $y \to 0$. The black solid, blue dashed, and red dotted curves correspond to decays into vector, fermion, and scalar final states, respectively. Projected sensitivities from different GW detection experiments are also shown in orange (adapted from Refs.~\cite{Ringwald:2020ist, Ringwald:2022xif}). The gray dashed diagonal lines are CMB bounds on $\DNeff$ from Planck, COrE/Euclid, and a hypothetical CVL experiment, respectively.}
    \label{fig:GWs}
\end{figure} 
Finally, we compute the dimensionless strain defined as~\cite{Maggiore:1999vm}
\begin{equation}
    h_c(f) = \frac{1}{f}\, \sqrt{\frac{3\, H_0^2\, \oGW(f)}{2 \pi^2}} = 1.26 \times 10^{-18}\, \left(\frac{\rm{Hz}}{f}\right) \sqrt{h^2\, \oGW(f)}\,,  
\end{equation}
where $H_0 \equiv H(T_0) \simeq 1.44 \times 10^{-42}$~GeV is the present-day Hubble parameter, and $h = 0.674$~\cite{Planck:2018vyg}.
In Fig.~\ref{fig:GWs} we show the dimensionless strain $h_c$ as a function of the GWs frequency $f$, for two benchmark points: \ding{192} $M = M_P/10$ and $\Trh = 5.5 \times 10^{15}$~GeV, and \ding{193} $M = M_P/10^3$ and $\Trh = M_P/(2 \times 10^4)$. In the same plane, we project the limits from several proposed GW detectors, for example, LISA~\cite{2017arXiv170200786A}, the Einstein Telescope (ET)~\cite{Punturo:2010zz, Hild:2010id, Sathyaprakash:2012jk, Maggiore:2019uih},  the Big Bang Observer (BBO)~\cite{Crowder:2005nr, Corbin:2005ny, Harry:2006fi}, ultimate DECIGO (uDECIGO)~\cite{Seto:2001qf, Kudoh:2005as}, GW-electromagnetic wave conversion in the vacuum (solid) and in a Gaussian beam (GB) (dotted)~\cite{Li:2003tv, Ringwald:2020ist}, resonant cavities~\cite{Berlin:2021txa, Herman:2022fau}, and the International Axion Observatory (IAXO)~\cite{Armengaud:2014gea, IAXO:2019mpb}. We have projected the $\DNeff$ bounds from Planck, COrE/Euclid and CVL from Fig.~\ref{fig:DNeff} as a bound on the GW strain using~\cite{Maggiore:1999vm} 
\begin{equation}
    \int d\left(\ln f\right) \oGW(f)\, h^2 \leq 5.6 \times 10^{-6}\, \DNeff\,.     
\end{equation}
These are shown by the diagonal straight gray lines. As can be inferred from Eq.~\eqref{eq:f}, a larger $M/\Trh$ ratio corresponds to a higher frequency, which is also reflected in Fig.~\ref{fig:GWs} curves. As we see, only microwave cavity detectors are capable of probing the high-frequency regime of the GW spectrum. Detectors such as uDECIGO, on the other hand, might be able to reach the lower frequency part of the spectrum. 

\section{Conclusions} \label{sec:concl}
Inflaton 3-body decay is a source of the stochastic gravitational wave (GW) background, due to the inexorable graviton Bremsstrahlung. In this work, we have revisited such three-body decay rates, considering the perturbative coupling of the inflaton with a pair of massive spin-0 bosons, spin-1/2 fermions, and spin-1 vector bosons, along with the radiative emission of a massless graviton from either the initial or the final states. We have found that the previously reported results show discrepancies with our findings in all three cases. To make our claim more robust, we employed two distinct procedures in calculating the graviton polarization and found that they agree with each other. 
With this improvement over existing results, we then numerically calculated the contribution of the GW energy density to the number of degrees of freedom around the time of BBN and CMB, typically encoded in $\DNeff$. We have taken care of the evolution of the energy densities of inflaton, radiation, and GW by solving a set of coupled Boltzmann equations. Due to Planck-scale suppression from minimal gravitational coupling, the GW energy density from inflaton Bremsstrahlung stays well below the CMB bounds on $\DNeff$, regardless of the spin of the final-state particles. As the spectrum of GW peaks in the GHz to THz ballpark, this primordial GW signature remains beyond the reach of most detector facilities; however, it may leave a footprint in resonant cavity detectors or even in upcoming space-based GW detectors.

\acknowledgments
The authors thank Manuel Drees, Yann Mambrini, Simon Cléry, and Marco Drewes for useful discussions, Rome Samanta for providing the experimental limits, and also Yong Tang and Da Huang for helpful communication. We thank Wenyuan Ai for pointing out typos in Eqs.~\eqref{eq:B11} and~\eqref{eq:B25}. We also thank Simon Cléry, Alejandro Ibarra, and Juntaro Wada for spotting two typos in Eq.~\eqref{eq:typo}. NB received funding from the Spanish FEDER / MCIU-AEI under the grant FPA2017-84543-P.  YX has received support from the Cluster of Excellence ``Precision Physics, Fundamental Interactions, and Structure of Matter'' (PRISMA$^+$ EXC 2118/1) funded by the Deutsche Forschungsgemeinschaft (DFG, German Research Foundation) within the German Excellence Strategy (Project No. 39083149). OZ has received funding from the Ministerio de Ciencia, Tecnología e Innovación (MinCiencias - Colombia) through Grants 82315-2021-1080 and 80740-492-2021, and has been partially supported by Sostenibilidad-UdeA and the UdeA/CODI Grant 2020-33177.

\appendix
\section{Feynman Rules for Relevant Vertices} \label{sec:app-vertex}
Here, we focus on a massless spin-2 graviton field, whose polarization tensor $\epsilon^{\mu\nu}$ has to satisfy the following conditions~\cite{Gross:1968in, Gleisberg:2003ue}
\begin{align}
    \epsilon^{i\, \mu\nu} = \epsilon^{i\, \nu\mu} &\qquad \text{symmetric}, \label{eq:conditon1}\\
    \omega_\mu\, \epsilon^{i\, \mu\nu} = 0 &\qquad \text{transverse},\\
    \eta_{\mu\nu}\, \epsilon^{i\, \mu\nu} = 0 &\qquad \text{traceless}, \label{eq:traceless}\\
    \epsilon^{i\, \mu\nu}\, \epsilon^{j\, \star}_{\mu\nu} = \delta^{ij} &\qquad \text{orthonormal},\label{eq:conditon4}
\end{align}
for $i,\, j = 1,\, 2$ being the polarization indices and $\omega$ the graviton four momentum. The polarization sum for the massless graviton is~\cite{deAquino:2011ix}
\begin{equation}\label{eq:tensor_pol_sum}
    \sum_\text{pol} \epsilon^{\star\mu\nu} \epsilon^{\alpha\beta} = \frac12 \left(\hat{\eta}^{\mu\alpha} \hat{\eta}^{\nu\beta} + \hat{\eta}^{\mu\beta} \hat{\eta}^{\nu\alpha} - \hat{\eta}^{\mu\nu} \hat{\eta}^{\alpha \beta}\right),
\end{equation}
with
\begin{equation}
    \hat{\eta}_{\mu \nu} \equiv \eta_{\mu \nu} - \frac{\omega_\mu \bar{\omega}_\nu +\bar{\omega}_\mu \omega_\nu}{\omega\cdot \bar{\omega}}\,,
\end{equation}
where $\omega = (E_\omega, \vec{\omega})$ and $\bar{\omega} = (E_\omega,- \vec{\omega})$. For a massless graviton, we have $\omega\cdot \bar{\omega} = E_\omega^2+\vec{\omega}^2 = 2E_\omega^2$.
The polarization sum in Eq.~\eqref{eq:tensor_pol_sum} indeed preserves the symmetric, transverse, traceless, and orthonormal conditions.\footnote{We note that the naive polarization sum $\sum_\text{pol} \epsilon^{\star\mu\nu} \epsilon^{\alpha\beta} = \frac12 \left(\eta^{\mu\alpha} \eta^{\nu\beta} + \eta^{\mu\beta} \eta^{\nu\alpha} -\eta^{\mu\nu} \eta^{\alpha \beta}\right)$ violates transverse, traceless, and orthonormal conditions and therefore should not be used.}
Note that due to the van Dam-Veltman discontinuity~\cite{vanDam:1970vg, deAquino:2011ix}, one cannot obtain the massless graviton propagator from the massive one simply by taking the limit of the graviton mass $m_{h_{\mu \nu}} \to 0$.

\begin{figure}[t!]
    \def\sepf{0.51}
    \centering
    \includegraphics[scale=0.14]{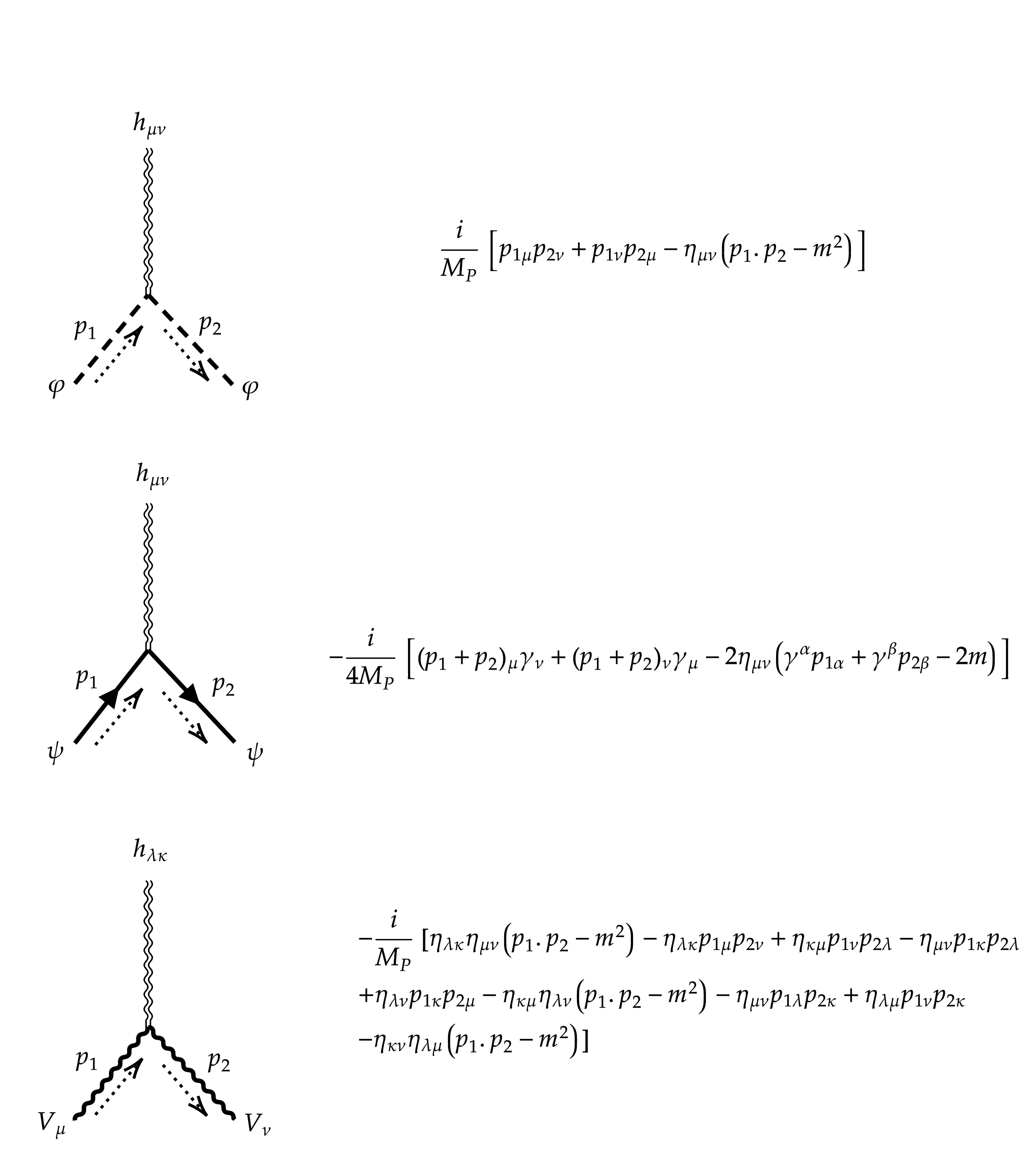}
    \caption{Relevant graviton-matter vertices for scalar $(\varphi)$, fermion $(\psi)$ and vector boson $(V)$, from top to bottom, following Ref.~\cite{Choi:1994ax}.}
    \label{fig:vertex}
\end{figure}

From the interaction Lagrangian in Eq.~\eqref{eq:int2} the relevant Feynman rules can be extracted, and we tabulate them in Fig.~\ref{fig:vertex}.

\section{Calculation of the Decay Widths} \label{sec:app-decay}
In this section, we present details of the computation of the differential rates for the inflaton decay into three-body final states, including a graviton.

\subsection{Scalar Case} \label{sec:app-boson}
To calculate the differential decay rate and cross-check the results, we present two different strategies based on: $i)$ an explicit construction for the graviton polarization tensor and $ii)$ polarization sum for the massless graviton.

\subsubsection{Polarization Tensor in Explicit Form}
Without losing generality, we choose a coordinate system in which the graviton moves along the $x$ direction, and hence the four-momentum of the graviton is $\omega = (E_\omega, \omega_x, 0, 0)$, where $\omega^2 = 0$, and then $\omega_x = E_\omega$. The four-momentum of the inflaton and its other two decay products are $l = (M,0,0,0)$, $p =(E_p, p_x, p_y, p_z)$, and $q = (M- E_p- E_\omega, -p_x-\omega_x, -p_y, -p_z)$, respectively.

The two polarization tensors that meet traceless, transverse, symmetric and orthonormal conditions described in Eqs.~\eqref{eq:conditon1} to~\eqref{eq:conditon4} can be explicitly written as~\cite{vanDam:1970vg}
\begin{equation}\label{eq:poltensor}
    \epsilon^1_{\mu\nu}=\frac{1}{\sqrt{2}}
    \begin{bmatrix}
        0 & 0 & 0& 0\\
        0 & 0 & 0& 0\\
        0 & 0 & 1& 0\\
        0 & 0 & 0& -1
    \end{bmatrix}
    \qquad \text{and} \qquad
    \epsilon^2_{\mu\nu}=\frac{1}{\sqrt{2}}
    \begin{bmatrix}
        0 & 0 & 0& 0\\
        0 & 0 & 0& 0\\
        0 & 0 & 0& 1\\
        0 & 0 & 1& 0
    \end{bmatrix}.
\end{equation}
Using the Feynman rules of Fig.~\ref{fig:vertex}, the amplitudes for the 3-body decays shown in Fig.~\ref{fig:feyn} are
\begin{align}\label{eq:M1}
    i\mathcal{M}_1 &= \frac{-i\, \mu}{M_P}\, \frac{l_\mu\, l_\nu\, \epsilon_i^{\star\mu\nu}}{M\, E_\omega} = 0\,, \\
    i\mathcal{M}_2 &= \frac{i\, \mu}{M_P}\, \frac{p_\mu\, p_\nu\, \epsilon_j^{\star\mu\nu}}{p \cdot \omega}\,, \\
    i\mathcal{M}_3 &= \frac{i\, \mu}{M_P}\, \frac{q_\mu\, q_\nu\, \epsilon_k^{\star\mu\nu}}{M\, E_\omega - p\, \cdot \omega}\,,\\
    i\mathcal{M}_4 &\propto \eta_{\mu \nu} \epsilon^{\mu \nu} =0\,,
\end{align}
where $\mathcal{M}_{1,2}$ corresponds to the diagrams in the upper left and upper right panels of Fig.~\ref{fig:feyn}, while $\mathcal{M}_{3,4}$ correspond to the lower left and lower right panels, respectively. Using Eq.~\eqref{eq:poltensor}, we notice that $\mathcal{M}_1=0$ as the decay takes place in the rest frame of the inflaton with $l = (M,0,0,0)$, while
\begin{equation}
    \sum_\text{pol} |\mathcal{M}_2|^2 = \frac{\mu^2}{(p\cdot \omega)^2 M_P^2} \sum_j (p_\mu p_\nu \epsilon_j^{\star \mu \nu}) (p_\mu p_\nu \epsilon_j^{\mu \nu})
    = \frac{\mu^2}{2(p\cdot \omega)^2M_P^2}  \left [ (p_y^2-p_z^2)^2  + 4p^2_y p^2_z \right].
\end{equation}
Similarly,
\begin{align}
    \sum_\text{pol} |\mathcal{M}_3|^2& = \frac{\mu^2}{2(M\, E_\omega - p \cdot \omega)^2\,M_P^2}  \left [ ((-p_y)^2 -(-p_z)^2)^2 +4 p_y^2 p_z^2 \right]\nonumber \\
    & = \frac{\mu^2}{2(M\, E_\omega - p \cdot \omega)^2\,M_P^2}  \left [ (p_y^2 -p_z^2)^2 +4 p_y^2 p_z^2 \right],
\end{align}
and the cross-term turns out to be
\begin{equation}
    \sum_\text{pol} \left(\mathcal{M}_2 \mathcal{M}^{\star}_3 \right) = \frac{\mu^2}{2(M\, E_\omega - p \cdot \omega) (p \cdot \omega) M_P^2}  \left[\left(p_y^2 - p_z^2\right)^2 + 4\, p_y^2\, p_z^2\right].
\end{equation}
Note that $\sum_\text{pol} \left(\mathcal{M}_1 \mathcal{M}^{\star}_2\right)=0$ and $ \sum_\text{pol} \left(\mathcal{M}_1 \mathcal{M}^{\star}_3\right)=0$ as $\mathcal{M}_1=0$.
The total squared amplitude is then
\begin{equation}
    \sum_\text{pol} |\mathcal{M}|^2 =\frac{\mu^2}{2 M_P^2} \left[\frac{1}{p \cdot \omega} + \frac{1}{M\, E_\omega - p \cdot \omega}\right]^2 \left(p_y^2 +p_z^2\right)^2.
\end{equation}
Since $ p \cdot \omega = E_p E_\omega - p_x E_\omega $, we have
\begin{equation}
    p_x = \frac{ E_p E_\omega -p \cdot \omega  }{E_\omega}\,, 
\end{equation}
which together with $m^2 \equiv p^2 = E_p^2 -(p_x^2 + p_y^2 + p_z^2)$ implies
\begin{align}\label{eq:B11}
   & \sum_\text{pol} |\mathcal{M}|^2 = \frac{\mu^2}{2 M_P^2} \left[\frac{1}{p \cdot \omega} + \frac{1}{M\, E_\omega - p \cdot \omega}\right]^2 \left[2\, \frac{E_p}{E_\omega}\, p \cdot \omega - \left(\frac{p \cdot \omega}{E_\omega}\right)^2 - m^2\right]^2 \nonumber \\
   &= \frac{\mu^2 \left[4 E_\omega^2 m^2 -8 E_p E_\omega M (E_\omega +E_p) + 4 M^2 (E_p^2 + 3 E_p E_\omega  +  E_\omega^2) - 4 (E_\omega +E_p)M^3 + M^4)\right]^2}{2\,M_P^2\, E_\omega^2\, M^2\, (M - 2\, E_p)^2 \left[M - 2\, (E_p +E_\omega)\right]^2}.
\end{align}
Finally, utilizing 
\begin{equation}
    \frac{d\Gamma}{dE_\omega} = \frac{1}{(2\pi)^3}\, \frac{1}{8\, M} \int_{E_{p,\text{min}}}^{E_{p,\text{max}}} dE_p\, |\mathcal{M}|^2,
\end{equation}
with
\begin{align}
    E_{p,\text{max}} &=\frac{1}{2} \left[M -E_\omega +E_\omega \sqrt{\frac{M^2 - 2 ME_\omega -4 m^2}{M(M-2E_\omega)}} \right],\\
    E_{p,\text{min}} &= \frac{1}{2} \left[M -E_\omega-E_\omega \sqrt{\frac{M^2 - 2 ME_\omega -4 m^2}{M(M-2E_\omega)}} \right],
\end{align}
one has the total differential cross-section as
\begin{equation} \label{eq:dGdE1}
   \frac{d\Gamma_0^{(1)}}{dE_\omega} = \frac{2}{64\, \pi^3} \left(\frac{\mu}{M_P}\right)^2\left[\frac{(2x-1)\, (2x -2y^2-1)}{4x\, \alpha^{-1}} + \frac{ y^2(y^2 + 2x - 1)}{x} \log\left(\frac{1 + \alpha}{1 - \alpha}\right)\right]
\end{equation}
for inflaton decays to complex scalars. Note that the extra factor 2 comes from two possible decay channels of the inflaton.

\subsubsection{Polarization Sum}
We now employ the second formalism of the calculation,
namely the tensor polarization sum formalism, as mentioned in Eq.~\eqref{eq:tensor_pol_sum}. The squared amplitudes are
\begin{align} \label{eq:M1square_scalar}
    \sum_\text{pol} |\mathcal{M}_1|^2 &= \frac{ \mu^2}{M^2 E_\omega^2\,M_P^2} l_{\mu} l_{\nu} l_{\alpha} l_{\beta}  \sum_\text{pol} \epsilon^{\star\mu\nu} \epsilon^{\alpha\beta} \nonumber \\  \nonumber
    & = \frac{ \mu^2}{2M^2 E_\omega^2\,M_P^2} l_{\mu} l_{\nu} l_{\alpha} l_{\beta}    \left(\hat{\eta}^{\mu\alpha} \hat{\eta}^{\nu\beta} + \hat{\eta}^{\mu\beta} \hat{\eta}^{\nu\alpha} - \hat{\eta}^{\mu\nu} \hat{\eta}^{\alpha \beta}\right)\\ \nonumber
    & = \frac{ \mu^2}{2M^2 E_\omega^2\,M_P^2} \left[l^2 l^2 -4l^2\frac{(l\cdot \omega) (l\cdot\bar{\omega})}{2E_\omega^2} + 4 \frac{(l\cdot \omega)^2 (l\cdot\bar{\omega})^2}{4E_\omega^4}\right]\nonumber \\
    & = \frac{ \mu^2}{2\,M^2 E_\omega^2\, M_P^2} \left[l^2 - \frac{(l\cdot \omega) (l\cdot\bar{\omega})}{E_\omega^2}\right]^2 = \frac{\mu^2}{2M^2 E_\omega^2\,M_P^2} \left(M^2 -M^2 \right) = 0\,,
\end{align}
\begin{align}
    \sum_\text{pol} |\mathcal{M}_2|^2 &=\frac{\mu^2}{M_P^2}\, \frac{p_\mu\, p_\nu\, p_\alpha\, p_\beta\,}{(p \cdot \omega)^2}\, \sum_\text{pol} \epsilon^{\star\mu\nu} \epsilon^{\alpha\beta} = \frac{\mu^2}{2 (p \cdot \omega)^2\,M_P^2} \left[p^2 - \frac{(p\cdot \omega) (p\cdot\bar{\omega})}{E_\omega^2}\right]^2 \nonumber \\
    & = \frac{ \mu^2}{2 (p \cdot \omega)^2\,M_P^2} \left[m^2 - \frac{(p\cdot \omega) (p\cdot\bar{\omega})}{E_\omega^2}\right]^2,
\end{align}
\begin{align}
    \sum_\text{pol} |\mathcal{M}_3|^2 
    & = \frac{ \mu^2}{2 (M\, E_\omega - p \cdot \omega)^2\,M_P^2} \left[q^2 - \frac{(q\cdot \omega) (q\cdot\bar{\omega})}{E_\omega^2}\right]^2 \nonumber \\
    & = \frac{ \mu^2}{2 (M\, E_\omega - p \cdot \omega)^2\,M_P^2} \left[m^2 - \frac{(q\cdot \omega) (q\cdot\bar{\omega})}{E_\omega^2}\right]^2,
\end{align}
\begin{align}
    \sum_\text{pol} \left( \mathcal{M}_2 \mathcal{M}^{\star}_3 \right)
    & = \frac{\mu^2}{ (p\cdot \omega) (M\, E_\omega - p \cdot \omega) \,M_P^2} p_{\mu} p_{\nu} q_{\alpha} q_{\beta}  \sum_\text{pol} \epsilon^{\star\mu\nu} \epsilon^{\alpha\beta}\nonumber \\
    & = \frac{ \mu^2}{2 (p\cdot \omega) (M\, E_\omega - p \cdot \omega) M_P^2}\, \Bigg\{ 2\left[(p\cdot q) - \frac{(p\cdot \omega) (q\cdot \bar{\omega}) + (p\cdot \bar{\omega}) (q\cdot \omega) }{2E_\omega^2}\right]^2 \nonumber \\
    & \qquad - \left[p^2 q^2 -\frac{p^2(q\cdot \omega)(q\cdot \bar{\omega}) +q^2(p\cdot \omega)(p\cdot \bar{\omega}) }{E_\omega^2} +\frac{(p\cdot \omega)(p\cdot \bar{\omega})(q\cdot \omega)(q\cdot \bar{\omega})}{E_\omega^4}\right]  \Bigg\} \nonumber \\
    & = \frac{ \mu^2}{2 (p\cdot \omega) (M\, E_\omega - p \cdot \omega) M_P^2}\, \Bigg\{ 2\left[(p\cdot q) - \frac{(p\cdot \omega) (q\cdot \bar{\omega}) + (p\cdot \bar{\omega}) (q\cdot \omega) }{2E_\omega^2}\right]^2 \nonumber\\
    & \qquad - \left[ p^2 -\frac{(q\cdot \omega)(q\cdot \bar{\omega})}{E_\omega^2} \right] \left[ q^2 -\frac{(p\cdot \omega)(p\cdot \bar{\omega})}{E_\omega^2} \right] \Bigg\}.
\end{align}
Note that Eq.~\eqref{eq:M1square_scalar} implies $\mathcal{M}_1=0$, which agrees with Eq.~\eqref{eq:M1}. Therefore, the other two interference terms $\sum_\text{pol} \left(\mathcal{M}_1 \mathcal{M}^{\star}_2\right) = \sum_\text{pol} \left(\mathcal{M}_1 \mathcal{M}^{\star}_3\right) = 0$.
Using the four vectors, one ends up with
\begin{align}
    p \cdot \omega &= M \left(E_\omega + E_p - \frac12 M\right),\\
    p \cdot \bar{\omega} &= 2 E_p E_\omega - p\cdot \omega\,,\\
    q\cdot \omega &= ME_\omega - p\cdot \omega\,,\\
    q\cdot \bar{\omega} &= (l- p -\omega ) \bar{\omega} = M E_\omega - p \cdot \bar{\omega} - 2E_\omega^2 =  M E_\omega - 2 E_p E_\omega + p\cdot \omega - 2E_\omega^2\,,\\
    p \cdot q  &= \frac{1}{2} \left[ (p+q)^2 - 2m^2   \right]  = \frac{1}{2} \left[ (l-\omega)^2 - 2m^2   \right] = \frac{1}{2} \left[ M^2-2M E_\omega - 2m^2   \right].
\end{align}
With these relations, one obtains
\begin{equation}\label{eq:B25}
    \sum_\text{pol} |\mathcal{M}|^2 
    = \frac{ \mu^2  \left[4E_\omega^2 m^2 -8 E_p E_\omega M (E_\omega +E_p ) + 4M^2 (E_p^2 + 3 E_p E_\omega +  E_\omega^2  ) - 4 (E_\omega +E_p)M^3 + M^4)\right]^2 }{2 M_P^2 E_\omega^2 M^2(M-2E_p)^2 [M- 2 (E_p + E_\omega) ]^2},
\end{equation}
and further
\begin{equation}\label{eq:dGdE2}
       \frac{d\Gamma_0^{(1)}}{dE_\omega}  = \frac{\mu^2}{32\pi^3 M_P^2} \left[\frac{(2x-1)\, (2x -2y^2-1)}{4x\, \alpha^{-1}} + \frac{ y^2(y^2 + 2x - 1)}{x} \log\left(\frac{1 + \alpha}{1 - \alpha}\right)\right],
\end{equation}
which agrees with Eq.~\eqref{eq:dGdE1}.

\subsection{Fermionic Case} \label{sec:app-ferm}
\begin{figure}[t!]
    \def\sepf{0.51}
	\centering
    \includegraphics[scale=0.26]{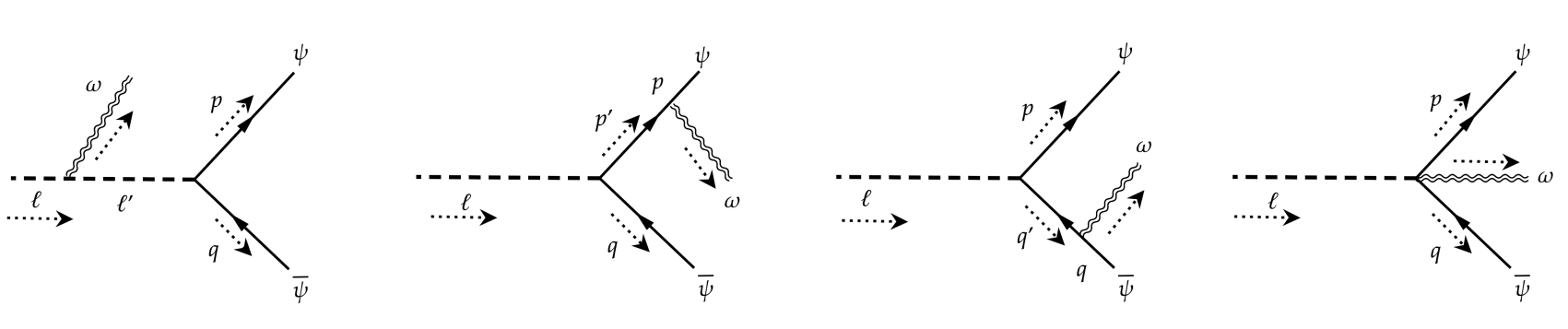}
    \caption{Feynman graph for inflaton decay into a pair of Dirac fermions with graviton emission.}
\label{fig:feyn_ferm}
\end{figure}

Using the list of Feynman rules in Fig.~\ref{fig:vertex}, the amplitudes for the inflaton decay into a Dirac fermion turn out to be
\begin{align}
    i \mathcal{M}_1 &= -i\, \frac{ y_\psi}{M_P}\, \frac{l_\mu\, l_\nu\, \epsilon^{\star\mu\nu}}{M\, E_\omega} \bar{u}(p) v(q)\,, \\
    i \mathcal{M}_2 &= \frac{i y_\psi}{2 p\cdot \omega M_P} \left[\bar{u}(p) (p_\mu \gamma_\nu) (\slashed{l}+ 2 m) v(q)\right] \epsilon^{*\mu \nu}, \\
    i \mathcal{M}_3 &= \frac{i y_\psi}{2 (M E_\omega - p\cdot \omega) M_P} \left[\bar{u}(p) (\slashed{l}- 2m)) (q_\mu \gamma_\nu)  v(q)\right] \epsilon^{*\mu \nu},\\
    i \mathcal{M}_4 &\propto \eta_{\mu \nu}\epsilon^{*\mu \nu} = 0\,,
\end{align}
where $\mathcal{M}_1, \mathcal{M}_2, \mathcal{M}_3~\text{and}~\mathcal{M}_4$ corresponds to the diagrams from left to right in Fig.~\ref{fig:feyn_ferm}.
Note that
\begin{align}
     \sum_{\text{spin, pol}}   |\mathcal{M}_1|^2 &= \frac{y_\psi^2}{M^2 E_\omega^2\,M_P^2} l_{\mu} l_{\nu} l_{\alpha} l_{\beta}  \sum_\text{pol} \epsilon^{\star\mu\nu} \epsilon^{\alpha\beta}  \times \text{Tr}\left[ (\slashed{q} -m) (\slashed{p} +m) \right]\,\nonumber\\
     & = \eqref{eq:M1square_scalar}\times \frac{y_\psi^2}{\mu^2} \times \text{Tr}\left[ (\slashed{q} -m) (\slashed{p} +m) \right]=0\,,
\end{align}
which again implies that $\mathcal{M}_1=0$. In fact, since $\mathcal{M}_1 \propto $ Eq.~\eqref{eq:M1}, one immediately finds that it vanishes. 
The other squared amplitudes are given by:
\begin{align}
\sum_{\text{spin, pol}}   |\mathcal{M}_2|^2 &= \frac{y_\psi^2}{(2 p\cdot \omega)^2\,M_P^2} \sum_\text{pol} \epsilon^{\star\mu\nu} \epsilon^{\alpha\beta} \nonumber\\
&\times \text{Tr}\left[ p_{\mu} \gamma_\nu (\slashed{l}+ 2 m) (\slashed{q} -m) (\slashed{l}+ 2 m) \gamma_\alpha p_{\beta}  (\slashed{p} +m) \right],\\
\sum_{\text{spin, pol}}   |\mathcal{M}_3|^2 &= \frac{y_\psi^2}{4 (M E_\omega - p\cdot \omega)^2\,M_P^2} \sum_\text{pol} \epsilon^{\star\mu\nu} \epsilon^{\alpha\beta}  \nonumber\\
&\times \text{Tr}\left[(\slashed{l}-2 m)  q_{\mu} \gamma_\nu  (\slashed{q} -m) \gamma_\alpha q_{\beta}  (\slashed{l}- 2 m) (\slashed{p} +m) \right],\\
\sum_{\text{spin, pol}}   (\mathcal{M}_2 \mathcal{M}_3^{*} ) &= \frac{y_\psi^2}{4 p\cdot \omega (M E_\omega - p\cdot \omega) \,M_P^2} \sum_\text{pol} \epsilon^{\star\mu\nu} \epsilon^{\alpha\beta}  \nonumber\\
&\times \text{Tr}\left[p_{\mu} \gamma_\nu (\slashed{l}+ 2 m) (\slashed{q} -m)\gamma_\alpha q_{\beta}  (\slashed{l}- 2 m) (\slashed{p} +m)  \right].
\end{align}
Similarly as before, the other interference terms $\sum_\text{pol} \left(\mathcal{M}_1 \mathcal{M}^{\star}_2\right) = \sum_\text{pol} \left(\mathcal{M}_1 \mathcal{M}^{\star}_3\right) = 0$.
The total matrix element squared turns out to be 
\begin{align} \label{eq:typo}
    & \sum_{\text{spin, pol}}|\mathcal{M}|^2 = y_{\psi}^2\frac{M(M-2 E_p) (M-2 E_\omega) ( 2 (E_p +E_\omega ) -M) -4 E_\omega^2 m^2}{ E_\omega^2 M^2 (M-2E_p)^2 (M- 2(E_p + E_\omega))^2 M_P^2} \nonumber\\ 
    & \qquad\quad \times  \Big\{ 4 M m^2 \left[ M (4 E_p^2 +12 E_p E_\omega + 3 E_\omega^2) - 4 M^2 (E_p + E_\omega) -8 E_p E_\omega  (E_p + E_\omega)  + M^3\right]\nonumber \\ 
    & \qquad\quad - M^2(M-2E_p) (2 E_\omega^2 - 2E_\omega M + M^2) \left[M - 2(E_p + E_\omega) \right] + 16 E_\omega^2 m^4  \Big\},
\end{align}
with which we find 
\begin{align} \label{eq:Ferdiff}
    \frac{d\Gamma_{1/2}^{(1)}}{dE_\omega} = \frac{ y_\psi^2 M^2}{64\,\pi^3 \, M_P^2} &\Bigg[\frac{(1-2x)}{x\alpha^{-1}} \left[8xy^2 + 2(x-1)x - 8y^4 - 2y^2 +1\right]\nonumber\\
    & + \frac{4y^2 \left[(5-8x)y^2 -(x-1)^2 -4y^4 \right] }{x}\log\left(\frac{1 + \alpha}{1 - \alpha}\right)\Bigg].
\end{align}

\subsection{Vector Case} \label{sec:app-vector}
\begin{figure}[t!]
    \centering
    \includegraphics[scale=0.25]{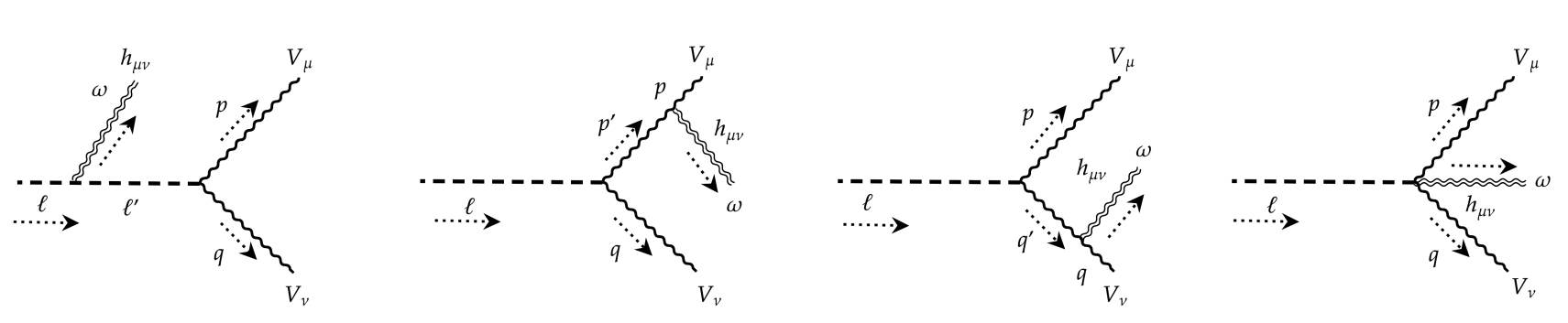}
    \caption{Feynman diagrams for inflaton decay into a pair of massive vectors with graviton emission.}
    \label{fig:feyn_vec}
\end{figure}

For the amplitudes of inflaton decay into massive spin-1 final states, we obtain
\begin{equation}
    i \mathcal{M}_1 = -i\, \frac{2\, g_V}{ M_P}\, \frac{l_\mu\, l_\nu\, \epsilon^{\star\mu\nu}}{M\, E_\omega} \eta_{\mu^{\prime} \nu^{\prime} } \varepsilon^{*\mu^\prime} (p,\lambda) \varepsilon^{*\nu^\prime} (q, \lambda) \,,
\end{equation}
and similarly, we find
\begin{align}
   i \mathcal{M}_2 &= -i\frac{ g_V }{p\cdot \omega\, M_P} \,\epsilon^{*\mu \nu} \varepsilon^{* \mu^\prime} (p,\lambda) \varepsilon^{* \nu^\prime} (q, \lambda) \cdot \eta_{\sigma  \nu^{\prime} }   \left[ \eta^{\rho \sigma} - \frac{(p^{\prime \rho} p^{\prime \sigma})}{m^2} \right]\nonumber \\
    & \quad \times \left[\eta_{\mu \nu} \eta_{\rho\mu^\prime}\left(p^{\prime} \cdot p -m^2\right)-\eta_{\mu \nu} p^{\prime}_{ \mu^\prime} p_{\rho}\right. +\eta_{\nu \rho} p^\prime_{ \mu^\prime} p_{ \mu}-\eta_{\rho\mu^\prime} p^\prime_{\nu} p_{\mu}+\eta_{\mu \mu^\prime} p^\prime_{ \nu} p_{\rho} \nonumber\\
    & \quad -\eta_{\nu \rho} \eta_{\mu \mu^\prime}\left(p^\prime \cdot p-m^2\right)+\eta_{\nu \mu^\prime} p^\prime_{\mu} p_{\rho}  -\eta_{\rho\mu^\prime} p^\prime_{ \mu} p_{ \nu}+\eta_{\mu \rho} p^\prime_{\mu^\prime} p_{\nu} \left.-\eta_{\nu \mu^\prime} \eta_{\mu \rho}\left(p^\prime \cdot p -m^2\right)\right],
\end{align}
\begin{align}
    i\mathcal{M}_3 &= -i\frac{g_V }{(M E_\omega - p\cdot \omega)\,M_P}\, \epsilon^{*\mu \nu} \varepsilon^{* \mu^\prime} (p,\lambda) \varepsilon^{* \nu^\prime} (q, \lambda) \cdot \eta_{\sigma  \mu^{\prime} }  \left[ \eta^{\rho \sigma} - \frac{(q^{\prime \rho} q^{\prime \sigma})}{m^2} \right]\nonumber \\
    & \quad \times \left[\eta_{\mu \nu} \eta_{\rho\nu^{\prime}}\left(q^{\prime} \cdot q -m^2\right)-\eta_{\mu \nu} q^{\prime}_{ \nu^{\prime}} q_{\rho}\right. +\eta_{\nu \rho} q^\prime_{ \nu^{\prime}} q_{ \mu}-\eta_{\rho\nu^{\prime}} q^\prime_{\nu} q_{\mu}+\eta_{\mu \nu^{\prime}} q^\prime_{ \nu} q_{\rho} \nonumber\\
    & \quad -\eta_{\nu \rho} \eta_{\mu \nu^{\prime}}\left(q^\prime \cdot q-m^2\right)+\eta_{\nu \nu^{\prime}} q^\prime_{\mu} q_{\rho}  -\eta_{\rho\nu^{\prime}} q^\prime_{ \mu} q_{ \nu}+\eta_{\mu \rho} q^\prime_{\nu^{\prime}} q_{\nu} \left.-\eta_{\nu \nu^{\prime}} \eta_{\mu \rho}\left(q^\prime \cdot q -m^2\right)\right],
\end{align}
while
\begin{equation}
    \mathcal{M}_4 \propto \eta_{\mu \nu}\epsilon^{*\mu \nu} = 0\,,
\end{equation}
for left to right in Fig.~\ref{fig:feyn_vec}, respectively.
The total squared matrix element is given by 
\begin{equation}
    \sum_{\text{spin, pol}}|\mathcal{M}|^2 = \frac{g_V^2}{2E_\omega^4\, M^2\, m^4 (M-2E_p)^2 (M - 2E_p -2E_\omega)^2 M_P^2 } \times \sum_{k=0}^8\,\mathcal{A}_k\,E_p^k\,,
\end{equation}
where
\begin{align}
    \mathcal{A}_8 &= 256 M^4 (M-2 E_\omega)^2,\\
    \mathcal{A}_7 &= 1024 (E_\omega -M)M^4 (M-2E_\omega)^2,\\
    \mathcal{A}_6 &= 128 (2E_\omega -M) M^3 \Big[ -2 E_\omega (M-4 E_\omega) m^2 -M (M-2E_\omega) (12 E_\omega^2 -27 M E_\omega + 14 M^2) \Big],\\
    \mathcal{A}_5 &= 128 M^3 (2E_\omega^2 -3 M E_\omega + M^2) \nonumber\\
    &\qquad \times \Big[ -6 E_\omega (M-4 E_\omega) m^2 -M(M-2E_\omega) (8 E_\omega^2 -25 ME_\omega +1 4 M^2)\Big],\\
    \mathcal{A}_4 &= 16 M^3 \Big[4 E_\omega^2 (3M-4 E_\omega) m^4\nonumber\\
    &\qquad + 4 E_\omega (M-2 E_\omega) (-60 E_\omega^3 + 138 M E_\omega^2 -91 M^2 E_\omega + 15 M^3 )\,m^2 \nonumber\\
    &\qquad + M(M-2E_\omega)^2 (16 E_\omega^4 -136 M E_\omega^3 + 301 M^2 E_\omega^2 -250 M^3 E_\omega + 70 M^4)\Big],\\
    \mathcal{A}_3 &= -32 (E_\omega -M) M^3 \Big[4 E_\omega^2 (4 E_\omega -3 M) m^4 -4 E_\omega (2E_\omega -M)\nonumber\\
    &\qquad \times (20 E_\omega^3 -48 M E_\omega^2 + 31 M^2 E_\omega -5 M^3 )m^2 + M^2\,(M-2E_\omega)^2 \nonumber\\
    &\qquad \times(12 E_\omega^3 -45 M E_\omega^2+ 46 M^2E_\omega - 14 M^3 ) \Big],\\
    \mathcal{A}_2 &= 8M \Big[M^5 (M-2E_\omega)(M-E_\omega) (-26 E_\omega^3 + 68 M E_\omega^2 -55 M^2 E_\omega + 14 M^3)\nonumber\\
    &\hspace{-0.5cm} + 2 E_\omega M^2 (M-2E_\omega) (-56 E_\omega^5 + 264 ME_\omega^4 -458 M^2 E_\omega^3 + 360 M^3 E_\omega^2 -126 M^4 E_\omega + 15M^5) m^2 \nonumber\\
    &\hspace{-0.5cm} + 16 E_\omega^4 (3 M-4 E_\omega ) m^6 + 4 E_\omega^2 M (-24 E_\omega^4 + 4 M E_\omega^3 + 26 M^2 E_\omega^2 -30 M^3 E_\omega + 9 M^4) m^4\Big],\\
    \mathcal{A}_1 &= 8 (E_\omega -M) M \Big[(2M - 3E_\omega) (E_\omega -M) (2E_\omega-M )^3 M^6 \nonumber\\
    &\quad +2E_\omega  (2E_\omega  -3 M) (M-2E_\omega  )^2 (2 E_\omega^3 - 8 M E_\omega^2 + 6 M^2 E_\omega - M^3)m^2 M^2  \nonumber\\
    &\quad +4E_\omega^2 (-24E_\omega^4 + 12 M E_\omega^3 + 4 M^2 E_\omega^2 -10 M^3 E_\omega + 3 M^4) m^4 M + 16 E_\omega^4 (3M - 4 E_\omega)m^6\Big],\\
    \mathcal{A}_0 &= 192\,E_\omega^6\,m^8 + 32\,E_\omega^4\,M^2 (6\,E_\omega^2-10\,M\,E_\omega+ 3\,M^2)\,m^6 + (E_\omega-M)^2\,M^8\,(M-2E_\omega)^4\nonumber\\
    &-4\,E_\omega^2\,M^2\,(2\,E_\omega^2 -2\,M\,E_\omega + M^2) (16\,E_\omega^4 -32\,M\,E_\omega^3 + 6\,M^2\,E_\omega^2 + 10\,M^3\,E_\omega -3\,M^4)\,m^4 \nonumber\\
    & - 4\,E_\omega\,M^4\,(M-2E_\omega)^4\,(2\,E_\omega^3 -4\,M E_\omega^2 + 5\,M^2\,E_\omega-M^3)\,m^2,
\end{align}
with which one can show that the differential decay rate reads
\begin{align} \label{eq:Vecdiff}
    \frac{d\Gamma^{(1)}_1}{dE_\omega} &=\frac{g_V^2}{1920\,\pi^3\,x\,y^4 \,M_P^2} \nonumber \\
    &\Bigg[ 60y^2 \left( -(1-2x)^2(1+2x) + (5+4x( -3 +4x)) y^2 + 16(x-1)y^4 + 12y^6 \right) \log \left(\frac{1 + \alpha}{1 - \alpha}\right)   \nonumber\\
    &  + \alpha  \Big( 360 (1-2 x) y^6+4 (4 x (23 x-5)+15) y^4+2 (2 x-1) (28 x (14 x-5)+15) y^2  \nonumber\\
    & +(1-2 x)^2 (4 x(2 x-5)+15) \Big) \Bigg].
\end{align}

\bibliographystyle{JHEP}
\bibliography{biblio}

\providecommand{\href}[2]{#2}\begingroup\raggedright\begin{thebibliography}{10}

\bibitem{Starobinsky:1979ty}
A.A.~Starobinsky, \emph{{Spectrum of relict gravitational radiation and the
  early state of the universe}}, {\emph{JETP Lett.} {\bfseries 30} (1979) 682}.

\bibitem{Allen:1987bk}
B.~Allen, \emph{{The Stochastic Gravity Wave Background in Inflationary
  Universe Models}},
  \href{https://doi.org/10.1103/PhysRevD.37.2078}{\emph{Phys. Rev. D}
  {\bfseries 37} (1988) 2078}.

\bibitem{Sahni:1990tx}
V.~Sahni, \emph{{The Energy Density of Relic Gravity Waves From Inflation}},
  \href{https://doi.org/10.1103/PhysRevD.42.453}{\emph{Phys. Rev. D} {\bfseries
  42} (1990) 453}.

\bibitem{Turner:1993vb}
M.S.~Turner, M.J.~White and J.E.~Lidsey, \emph{{Tensor perturbations in
  inflationary models as a probe of cosmology}},
  \href{https://doi.org/10.1103/PhysRevD.48.4613}{\emph{Phys. Rev. D}
  {\bfseries 48} (1993) 4613}
  [\href{https://arxiv.org/abs/astro-ph/9306029}{{\ttfamily
  astro-ph/9306029}}].

\bibitem{Khlebnikov:1997di}
S.Y.~Khlebnikov and I.I.~Tkachev, \emph{{Relic gravitational waves produced
  after preheating}},
  \href{https://doi.org/10.1103/PhysRevD.56.653}{\emph{Phys. Rev. D} {\bfseries
  56} (1997) 653} [\href{https://arxiv.org/abs/hep-ph/9701423}{{\ttfamily
  hep-ph/9701423}}].

\bibitem{Easther:2006gt}
R.~Easther and E.A.~Lim, \emph{{Stochastic gravitational wave production after
  inflation}}, \href{https://doi.org/10.1088/1475-7516/2006/04/010}{\emph{JCAP}
  {\bfseries 04} (2006) 010}
  [\href{https://arxiv.org/abs/astro-ph/0601617}{{\ttfamily
  astro-ph/0601617}}].

\bibitem{Dufaux:2007pt}
J.F.~Dufaux, A.~Bergman, G.N.~Felder, L.~Kofman and J.-P.~Uzan, \emph{{Theory
  and Numerics of Gravitational Waves from Preheating after Inflation}},
  \href{https://doi.org/10.1103/PhysRevD.76.123517}{\emph{Phys. Rev. D}
  {\bfseries 76} (2007) 123517}
  [\href{https://arxiv.org/abs/0707.0875}{{\ttfamily 0707.0875}}].

\bibitem{Bethke:2013aba}
L.~Bethke, D.G.~Figueroa and A.~Rajantie, \emph{{Anisotropies in the
  Gravitational Wave Background from Preheating}},
  \href{https://doi.org/10.1103/PhysRevLett.111.011301}{\emph{Phys. Rev. Lett.}
  {\bfseries 111} (2013) 011301}
  [\href{https://arxiv.org/abs/1304.2657}{{\ttfamily 1304.2657}}].

\bibitem{Figueroa:2017vfa}
D.G.~Figueroa and F.~Torrenti, \emph{{Gravitational wave production from
  preheating: parameter dependence}},
  \href{https://doi.org/10.1088/1475-7516/2017/10/057}{\emph{JCAP} {\bfseries
  10} (2017) 057} [\href{https://arxiv.org/abs/1707.04533}{{\ttfamily
  1707.04533}}].

\bibitem{Vilenkin:1981bx}
A.~Vilenkin, \emph{{Gravitational radiation from cosmic strings}},
  \href{https://doi.org/10.1016/0370-2693(81)91144-8}{\emph{Phys. Lett. B}
  {\bfseries 107} (1981) 47}.

\bibitem{Cui:2017ufi}
Y.~Cui, M.~Lewicki, D.E.~Morrissey and J.D.~Wells, \emph{{Cosmic Archaeology
  with Gravitational Waves from Cosmic Strings}},
  \href{https://doi.org/10.1103/PhysRevD.97.123505}{\emph{Phys. Rev. D}
  {\bfseries 97} (2018) 123505}
  [\href{https://arxiv.org/abs/1711.03104}{{\ttfamily 1711.03104}}].

\bibitem{Cui:2018rwi}
Y.~Cui, M.~Lewicki, D.E.~Morrissey and J.D.~Wells, \emph{{Probing the pre-BBN
  universe with gravitational waves from cosmic strings}},
  \href{https://doi.org/10.1007/JHEP01(2019)081}{\emph{JHEP} {\bfseries 01}
  (2019) 081} [\href{https://arxiv.org/abs/1808.08968}{{\ttfamily
  1808.08968}}].

\bibitem{Chang:2021afa}
C.-F.~Chang and Y.~Cui, \emph{{Gravitational waves from global cosmic strings
  and cosmic archaeology}},
  \href{https://doi.org/10.1007/JHEP03(2022)114}{\emph{JHEP} {\bfseries 03}
  (2022) 114} [\href{https://arxiv.org/abs/2106.09746}{{\ttfamily
  2106.09746}}].

\bibitem{Ghiglieri:2015nfa}
J.~Ghiglieri and M.~Laine, \emph{{Gravitational wave background from Standard
  Model physics: Qualitative features}},
  \href{https://doi.org/10.1088/1475-7516/2015/07/022}{\emph{JCAP} {\bfseries
  07} (2015) 022} [\href{https://arxiv.org/abs/1504.02569}{{\ttfamily
  1504.02569}}].

\bibitem{Ghiglieri:2020mhm}
J.~Ghiglieri, G.~Jackson, M.~Laine and Y.~Zhu, \emph{{Gravitational wave
  background from Standard Model physics: Complete leading order}},
  \href{https://doi.org/10.1007/JHEP07(2020)092}{\emph{JHEP} {\bfseries 07}
  (2020) 092} [\href{https://arxiv.org/abs/2004.11392}{{\ttfamily
  2004.11392}}].

\bibitem{Ghiglieri:2022rfp}
J.~Ghiglieri, J.~Sch\"utte-Engel and E.~Speranza, \emph{{Freezing-In
  Gravitational Waves}},  \href{https://arxiv.org/abs/2211.16513}{{\ttfamily
  2211.16513}}.

\bibitem{Nakayama:2018ptw}
K.~Nakayama and Y.~Tang, \emph{{Stochastic Gravitational Waves from Particle
  Origin}}, \href{https://doi.org/10.1016/j.physletb.2018.11.023}{\emph{Phys.
  Lett. B} {\bfseries 788} (2019) 341}
  [\href{https://arxiv.org/abs/1810.04975}{{\ttfamily 1810.04975}}].

\bibitem{Huang:2019lgd}
D.~Huang and L.~Yin, \emph{{Stochastic Gravitational Waves from Inflaton
  Decays}}, \href{https://doi.org/10.1103/PhysRevD.100.043538}{\emph{Phys. Rev.
  D} {\bfseries 100} (2019) 043538}
  [\href{https://arxiv.org/abs/1905.08510}{{\ttfamily 1905.08510}}].

\bibitem{Mambrini:2022uol}
Y.~Mambrini, K.A.~Olive and J.~Zheng, \emph{{Post-inflationary dark matter
  bremsstrahlung}},
  \href{https://doi.org/10.1088/1475-7516/2022/10/055}{\emph{JCAP} {\bfseries
  10} (2022) 055} [\href{https://arxiv.org/abs/2208.05859}{{\ttfamily
  2208.05859}}].

\bibitem{Weinberg:1965nx}
S.~Weinberg, \emph{{Infrared photons and gravitons}},
  \href{https://doi.org/10.1103/PhysRev.140.B516}{\emph{Phys. Rev.} {\bfseries
  140} (1965) B516}.

\bibitem{Dufaux:2006ee}
J.F.~Dufaux, G.N.~Felder, L.~Kofman, M.~Peloso and D.~Podolsky,
  \emph{{Preheating with trilinear interactions: Tachyonic resonance}},
  \href{https://doi.org/10.1088/1475-7516/2006/07/006}{\emph{JCAP} {\bfseries
  07} (2006) 006} [\href{https://arxiv.org/abs/hep-ph/0602144}{{\ttfamily
  hep-ph/0602144}}].

\bibitem{Ghoshal:2022kqp}
A.~Ghoshal, R.~Samanta and G.~White, \emph{{Bremsstrahlung High-frequency
  Gravitational Wave Signatures of High-scale Non-thermal Leptogenesis}},
  \href{https://arxiv.org/abs/2211.10433}{{\ttfamily 2211.10433}}.

\bibitem{Pagano:2015hma}
L.~Pagano, L.~Salvati and A.~Melchiorri, \emph{{New constraints on primordial
  gravitational waves from Planck 2015}},
  \href{https://doi.org/10.1016/j.physletb.2016.07.078}{\emph{Phys. Lett. B}
  {\bfseries 760} (2016) 823}
  [\href{https://arxiv.org/abs/1508.02393}{{\ttfamily 1508.02393}}].

\bibitem{Caprini:2018mtu}
C.~Caprini and D.G.~Figueroa, \emph{{Cosmological Backgrounds of Gravitational
  Waves}}, \href{https://doi.org/10.1088/1361-6382/aac608}{\emph{Class. Quant.
  Grav.} {\bfseries 35} (2018) 163001}
  [\href{https://arxiv.org/abs/1801.04268}{{\ttfamily 1801.04268}}].

\bibitem{Berlin:2021txa}
A.~Berlin, D.~Blas, R.~Tito~D'Agnolo, S.A.R.~Ellis, R.~Harnik, Y.~Kahn et~al.,
  \emph{{Detecting high-frequency gravitational waves with microwave
  cavities}}, \href{https://doi.org/10.1103/PhysRevD.105.116011}{\emph{Phys.
  Rev. D} {\bfseries 105} (2022) 116011}
  [\href{https://arxiv.org/abs/2112.11465}{{\ttfamily 2112.11465}}].

\bibitem{Herman:2022fau}
N.~Herman, L.~Lehoucq and A.~F\'{u}zfa, \emph{{Electromagnetic Antennas for the
  Resonant Detection of the Stochastic Gravitational Wave Background}},
  \href{https://arxiv.org/abs/2203.15668}{{\ttfamily 2203.15668}}.

\bibitem{Seto:2001qf}
N.~Seto, S.~Kawamura and T.~Nakamura, \emph{{Possibility of direct measurement
  of the acceleration of the universe using 0.1-Hz band laser interferometer
  gravitational wave antenna in space}},
  \href{https://doi.org/10.1103/PhysRevLett.87.221103}{\emph{Phys. Rev. Lett.}
  {\bfseries 87} (2001) 221103}
  [\href{https://arxiv.org/abs/astro-ph/0108011}{{\ttfamily
  astro-ph/0108011}}].

\bibitem{Kudoh:2005as}
H.~Kudoh, A.~Taruya, T.~Hiramatsu and Y.~Himemoto, \emph{{Detecting a
  gravitational-wave background with next-generation space interferometers}},
  \href{https://doi.org/10.1103/PhysRevD.73.064006}{\emph{Phys. Rev. D}
  {\bfseries 73} (2006) 064006}
  [\href{https://arxiv.org/abs/gr-qc/0511145}{{\ttfamily gr-qc/0511145}}].

\bibitem{Aggarwal:2020olq}
N.~Aggarwal et~al., \emph{{Challenges and opportunities of gravitational-wave
  searches at MHz to GHz frequencies}},
  \href{https://doi.org/10.1007/s41114-021-00032-5}{\emph{Living Rev. Rel.}
  {\bfseries 24} (2021) 4} [\href{https://arxiv.org/abs/2011.12414}{{\ttfamily
  2011.12414}}].

\bibitem{Choi:1994ax}
S.Y.~Choi, J.S.~Shim and H.S.~Song, \emph{{Factorization and polarization in
  linearized gravity}},
  \href{https://doi.org/10.1103/PhysRevD.51.2751}{\emph{Phys. Rev. D}
  {\bfseries 51} (1995) 2751}
  [\href{https://arxiv.org/abs/hep-th/9411092}{{\ttfamily hep-th/9411092}}].

\bibitem{Kofman:1997yn}
L.~Kofman, A.D.~Linde and A.A.~Starobinsky, \emph{{Towards the theory of
  reheating after inflation}},
  \href{https://doi.org/10.1103/PhysRevD.56.3258}{\emph{Phys. Rev. D}
  {\bfseries 56} (1997) 3258}
  [\href{https://arxiv.org/abs/hep-ph/9704452}{{\ttfamily hep-ph/9704452}}].

\bibitem{Drewes:2019rxn}
M.~Drewes, \emph{{Measuring the inflaton coupling in~the~CMB}},
  \href{https://doi.org/10.1088/1475-7516/2022/09/069}{\emph{JCAP} {\bfseries
  09} (2022) 069} [\href{https://arxiv.org/abs/1903.09599}{{\ttfamily
  1903.09599}}].

\bibitem{Drewes:2022nhu}
M.~Drewes and L.~Ming, \emph{{Connecting Cosmic Inflation to Particle Physics
  with LiteBIRD, CMB S4, EUCLID and SKA}},
  \href{https://arxiv.org/abs/2208.07609}{{\ttfamily 2208.07609}}.

\bibitem{Drees:2021wgd}
M.~Drees and Y.~Xu, \emph{{Small field polynomial inflation: reheating,
  radiative stability and lower bound}},
  \href{https://doi.org/10.1088/1475-7516/2021/09/012}{\emph{JCAP} {\bfseries
  09} (2021) 012} [\href{https://arxiv.org/abs/2104.03977}{{\ttfamily
  2104.03977}}].

\bibitem{Peloso:2000hy}
M.~Peloso and L.~Sorbo, \emph{{Preheating of massive fermions after inflation:
  Analytical results}},
  \href{https://doi.org/10.1088/1126-6708/2000/05/016}{\emph{JHEP} {\bfseries
  05} (2000) 016} [\href{https://arxiv.org/abs/hep-ph/0003045}{{\ttfamily
  hep-ph/0003045}}].

\bibitem{Felder:1998vq}
G.N.~Felder, L.~Kofman and A.D.~Linde, \emph{{Instant preheating}},
  \href{https://doi.org/10.1103/PhysRevD.59.123523}{\emph{Phys. Rev. D}
  {\bfseries 59} (1999) 123523}
  [\href{https://arxiv.org/abs/hep-ph/9812289}{{\ttfamily hep-ph/9812289}}].

\bibitem{Klose:2022knn}
P.~Klose, M.~Laine and S.~Procacci, \emph{{Gravitational wave background from
  non-Abelian reheating after axion-like inflation}},
  \href{https://doi.org/10.1088/1475-7516/2022/05/021}{\emph{JCAP} {\bfseries
  05} (2022) 021} [\href{https://arxiv.org/abs/2201.02317}{{\ttfamily
  2201.02317}}].

\bibitem{Sarkar:1995dd}
S.~Sarkar, \emph{{Big bang nucleosynthesis and physics beyond the standard
  model}}, \href{https://doi.org/10.1088/0034-4885/59/12/001}{\emph{Rept. Prog.
  Phys.} {\bfseries 59} (1996) 1493}
  [\href{https://arxiv.org/abs/hep-ph/9602260}{{\ttfamily hep-ph/9602260}}].

\bibitem{Kawasaki:2000en}
M.~Kawasaki, K.~Kohri and N.~Sugiyama, \emph{{MeV scale reheating temperature
  and thermalization of neutrino background}},
  \href{https://doi.org/10.1103/PhysRevD.62.023506}{\emph{Phys. Rev. D}
  {\bfseries 62} (2000) 023506}
  [\href{https://arxiv.org/abs/astro-ph/0002127}{{\ttfamily
  astro-ph/0002127}}].

\bibitem{Hannestad:2004px}
S.~Hannestad, \emph{{What is the lowest possible reheating temperature?}},
  \href{https://doi.org/10.1103/PhysRevD.70.043506}{\emph{Phys. Rev. D}
  {\bfseries 70} (2004) 043506}
  [\href{https://arxiv.org/abs/astro-ph/0403291}{{\ttfamily
  astro-ph/0403291}}].

\bibitem{DeBernardis:2008zz}
F.~De~Bernardis, L.~Pagano and A.~Melchiorri, \emph{{New constraints on the
  reheating temperature of the universe after WMAP-5}},
  \href{https://doi.org/10.1016/j.astropartphys.2008.09.005}{\emph{Astropart.
  Phys.} {\bfseries 30} (2008) 192}.

\bibitem{deSalas:2015glj}
P.F.~de~Salas, M.~Lattanzi, G.~Mangano, G.~Miele, S.~Pastor and O.~Pisanti,
  \emph{{Bounds on very low reheating scenarios after Planck}},
  \href{https://doi.org/10.1103/PhysRevD.92.123534}{\emph{Phys. Rev. D}
  {\bfseries 92} (2015) 123534}
  [\href{https://arxiv.org/abs/1511.00672}{{\ttfamily 1511.00672}}].

\bibitem{Dodelson:1992km}
S.~Dodelson and M.S.~Turner, \emph{{Nonequilibrium neutrino statistical
  mechanics in the expanding universe}},
  \href{https://doi.org/10.1103/PhysRevD.46.3372}{\emph{Phys. Rev. D}
  {\bfseries 46} (1992) 3372}.

\bibitem{Hannestad:1995rs}
S.~Hannestad and J.~Madsen, \emph{{Neutrino decoupling in the early universe}},
  \href{https://doi.org/10.1103/PhysRevD.52.1764}{\emph{Phys. Rev. D}
  {\bfseries 52} (1995) 1764}
  [\href{https://arxiv.org/abs/astro-ph/9506015}{{\ttfamily
  astro-ph/9506015}}].

\bibitem{Dolgov:1997mb}
A.D.~Dolgov, S.H.~Hansen and D.V.~Semikoz, \emph{{Nonequilibrium corrections to
  the spectra of massless neutrinos in the early universe}},
  \href{https://doi.org/10.1016/S0550-3213(97)00479-3}{\emph{Nucl. Phys. B}
  {\bfseries 503} (1997) 426}
  [\href{https://arxiv.org/abs/hep-ph/9703315}{{\ttfamily hep-ph/9703315}}].

\bibitem{Mangano:2005cc}
G.~Mangano, G.~Miele, S.~Pastor, T.~Pinto, O.~Pisanti and P.D.~Serpico,
  \emph{{Relic neutrino decoupling including flavor oscillations}},
  \href{https://doi.org/10.1016/j.nuclphysb.2005.09.041}{\emph{Nucl. Phys. B}
  {\bfseries 729} (2005) 221}
  [\href{https://arxiv.org/abs/hep-ph/0506164}{{\ttfamily hep-ph/0506164}}].

\bibitem{deSalas:2016ztq}
P.F.~de~Salas and S.~Pastor, \emph{{Relic neutrino decoupling with flavour
  oscillations revisited}},
  \href{https://doi.org/10.1088/1475-7516/2016/07/051}{\emph{JCAP} {\bfseries
  07} (2016) 051} [\href{https://arxiv.org/abs/1606.06986}{{\ttfamily
  1606.06986}}].

\bibitem{EscuderoAbenza:2020cmq}
M.~Escudero~Abenza, \emph{{Precision early universe thermodynamics made simple:
  $N_{\rm eff}$ and neutrino decoupling in the Standard Model and beyond}},
  \href{https://doi.org/10.1088/1475-7516/2020/05/048}{\emph{JCAP} {\bfseries
  05} (2020) 048} [\href{https://arxiv.org/abs/2001.04466}{{\ttfamily
  2001.04466}}].

\bibitem{Akita:2020szl}
K.~Akita and M.~Yamaguchi, \emph{{A precision calculation of relic neutrino
  decoupling}},
  \href{https://doi.org/10.1088/1475-7516/2020/08/012}{\emph{JCAP} {\bfseries
  08} (2020) 012} [\href{https://arxiv.org/abs/2005.07047}{{\ttfamily
  2005.07047}}].

\bibitem{Froustey:2020mcq}
J.~Froustey, C.~Pitrou and M.C.~Volpe, \emph{{Neutrino decoupling including
  flavour oscillations and primordial nucleosynthesis}},
  \href{https://doi.org/10.1088/1475-7516/2020/12/015}{\emph{JCAP} {\bfseries
  12} (2020) 015} [\href{https://arxiv.org/abs/2008.01074}{{\ttfamily
  2008.01074}}].

\bibitem{Bennett:2020zkv}
J.J.~Bennett, G.~Buldgen, P.F.~De~Salas, M.~Drewes, S.~Gariazzo, S.~Pastor
  et~al., \emph{{Towards a precision calculation of $N_{\rm eff}$ in the
  Standard Model II: Neutrino decoupling in the presence of flavour
  oscillations and finite-temperature QED}},
  \href{https://doi.org/10.1088/1475-7516/2021/04/073}{\emph{JCAP} {\bfseries
  04} (2021) 073} [\href{https://arxiv.org/abs/2012.02726}{{\ttfamily
  2012.02726}}].

\bibitem{Drees:2015exa}
M.~Drees, F.~Hajkarim and E.R.~Schmitz, \emph{{The Effects of QCD Equation of
  State on the Relic Density of WIMP Dark Matter}},
  \href{https://doi.org/10.1088/1475-7516/2015/06/025}{\emph{JCAP} {\bfseries
  06} (2015) 025} [\href{https://arxiv.org/abs/1503.03513}{{\ttfamily
  1503.03513}}].

\bibitem{Giudice:2000ex}
G.F.~Giudice, E.W.~Kolb and A.~Riotto, \emph{{Largest temperature of the
  radiation era and its cosmological implications}},
  \href{https://doi.org/10.1103/PhysRevD.64.023508}{\emph{Phys. Rev. D}
  {\bfseries 64} (2001) 023508}
  [\href{https://arxiv.org/abs/hep-ph/0005123}{{\ttfamily hep-ph/0005123}}].

\bibitem{Planck:2018vyg}
{\scshape Planck} collaboration, \emph{{Planck 2018 results. VI. Cosmological
  parameters}},
  \href{https://doi.org/10.1051/0004-6361/201833910}{\emph{Astron. Astrophys.}
  {\bfseries 641} (2020) A6}
  [\href{https://arxiv.org/abs/1807.06209}{{\ttfamily 1807.06209}}].

\bibitem{BICEP:2021xfz}
{\scshape BICEP, Keck} collaboration, \emph{{Improved Constraints on Primordial
  Gravitational Waves using Planck, WMAP, and BICEP/Keck Observations through
  the 2018 Observing Season}},
  \href{https://doi.org/10.1103/PhysRevLett.127.151301}{\emph{Phys. Rev. Lett.}
  {\bfseries 127} (2021) 151301}
  [\href{https://arxiv.org/abs/2110.00483}{{\ttfamily 2110.00483}}].

\bibitem{Yeh:2022heq}
T.-H.~Yeh, J.~Shelton, K.A.~Olive and B.D.~Fields, \emph{{Probing physics
  beyond the standard model: limits from BBN and the CMB independently and
  combined}}, \href{https://doi.org/10.1088/1475-7516/2022/10/046}{\emph{JCAP}
  {\bfseries 10} (2022) 046}
  [\href{https://arxiv.org/abs/2207.13133}{{\ttfamily 2207.13133}}].

\bibitem{Abazajian:2019eic}
K.~Abazajian et~al., \emph{{CMB-S4 Science Case, Reference Design, and Project
  Plan}},  \href{https://arxiv.org/abs/1907.04473}{{\ttfamily 1907.04473}}.

\bibitem{CMB-HD:2022bsz}
{\scshape CMB-HD} collaboration, \emph{{Snowmass2021 CMB-HD White Paper}},
  \href{https://arxiv.org/abs/2203.05728}{{\ttfamily 2203.05728}}.

\bibitem{COrE:2011bfs}
{\scshape COrE} collaboration, \emph{{COrE (Cosmic Origins Explorer) A White
  Paper}},  \href{https://arxiv.org/abs/1102.2181}{{\ttfamily 1102.2181}}.

\bibitem{EUCLID:2011zbd}
{\scshape EUCLID} collaboration, \emph{{Euclid Definition Study Report}},
  \href{https://arxiv.org/abs/1110.3193}{{\ttfamily 1110.3193}}.

\bibitem{Ben-Dayan:2019gll}
I.~Ben-Dayan, B.~Keating, D.~Leon and I.~Wolfson, \emph{{Constraints on scalar
  and tensor spectra from $N_{eff}$}},
  \href{https://doi.org/10.1088/1475-7516/2019/06/007}{\emph{JCAP} {\bfseries
  06} (2019) 007} [\href{https://arxiv.org/abs/1903.11843}{{\ttfamily
  1903.11843}}].

\bibitem{SPT-3G:2014dbx}
{\scshape SPT-3G} collaboration, \emph{{SPT-3G: A Next-Generation Cosmic
  Microwave Background Polarization Experiment on the South Pole Telescope}},
  \href{https://doi.org/10.1117/12.2057305}{\emph{Proc. SPIE Int. Soc. Opt.
  Eng.} {\bfseries 9153} (2014) 91531P}
  [\href{https://arxiv.org/abs/1407.2973}{{\ttfamily 1407.2973}}].

\bibitem{SimonsObservatory:2018koc}
{\scshape Simons Observatory} collaboration, \emph{{The Simons Observatory:
  Science goals and forecasts}},
  \href{https://doi.org/10.1088/1475-7516/2019/02/056}{\emph{JCAP} {\bfseries
  02} (2019) 056} [\href{https://arxiv.org/abs/1808.07445}{{\ttfamily
  1808.07445}}].

\bibitem{Ringwald:2020ist}
A.~Ringwald, J.~Sch\"utte-Engel and C.~Tamarit, \emph{{Gravitational Waves as a
  Big Bang Thermometer}},
  \href{https://doi.org/10.1088/1475-7516/2021/03/054}{\emph{JCAP} {\bfseries
  03} (2021) 054} [\href{https://arxiv.org/abs/2011.04731}{{\ttfamily
  2011.04731}}].

\bibitem{Ringwald:2022xif}
A.~Ringwald and C.~Tamarit, \emph{{Revealing the cosmic history with
  gravitational waves}},
  \href{https://doi.org/10.1103/PhysRevD.106.063027}{\emph{Phys. Rev. D}
  {\bfseries 106} (2022) 063027}
  [\href{https://arxiv.org/abs/2203.00621}{{\ttfamily 2203.00621}}].

\bibitem{Maggiore:1999vm}
M.~Maggiore, \emph{{Gravitational wave experiments and early universe
  cosmology}}, \href{https://doi.org/10.1016/S0370-1573(99)00102-7}{\emph{Phys.
  Rept.} {\bfseries 331} (2000) 283}
  [\href{https://arxiv.org/abs/gr-qc/9909001}{{\ttfamily gr-qc/9909001}}].

\bibitem{2017arXiv170200786A}
{\scshape LISA} collaboration, \emph{{Laser Interferometer Space Antenna}},
  {\emph{arXiv e-prints} (2017) arXiv:1702.00786}
  [\href{https://arxiv.org/abs/1702.00786}{{\ttfamily 1702.00786}}].

\bibitem{Punturo:2010zz}
M.~Punturo et~al., \emph{{The Einstein Telescope: A third-generation
  gravitational wave observatory}},
  \href{https://doi.org/10.1088/0264-9381/27/19/194002}{\emph{Class. Quant.
  Grav.} {\bfseries 27} (2010) 194002}.

\bibitem{Hild:2010id}
S.~Hild et~al., \emph{{Sensitivity Studies for Third-Generation Gravitational
  Wave Observatories}},
  \href{https://doi.org/10.1088/0264-9381/28/9/094013}{\emph{Class. Quant.
  Grav.} {\bfseries 28} (2011) 094013}
  [\href{https://arxiv.org/abs/1012.0908}{{\ttfamily 1012.0908}}].

\bibitem{Sathyaprakash:2012jk}
B.~Sathyaprakash et~al., \emph{{Scientific Objectives of Einstein Telescope}},
  \href{https://doi.org/10.1088/0264-9381/29/12/124013}{\emph{Class. Quant.
  Grav.} {\bfseries 29} (2012) 124013}
  [\href{https://arxiv.org/abs/1206.0331}{{\ttfamily 1206.0331}}].

\bibitem{Maggiore:2019uih}
M.~Maggiore et~al., \emph{{Science Case for the Einstein Telescope}},
  \href{https://doi.org/10.1088/1475-7516/2020/03/050}{\emph{JCAP} {\bfseries
  03} (2020) 050} [\href{https://arxiv.org/abs/1912.02622}{{\ttfamily
  1912.02622}}].

\bibitem{Crowder:2005nr}
J.~Crowder and N.J.~Cornish, \emph{{Beyond LISA: Exploring future gravitational
  wave missions}},
  \href{https://doi.org/10.1103/PhysRevD.72.083005}{\emph{Phys. Rev. D}
  {\bfseries 72} (2005) 083005}
  [\href{https://arxiv.org/abs/gr-qc/0506015}{{\ttfamily gr-qc/0506015}}].

\bibitem{Corbin:2005ny}
V.~Corbin and N.J.~Cornish, \emph{{Detecting the cosmic gravitational wave
  background with the big bang observer}},
  \href{https://doi.org/10.1088/0264-9381/23/7/014}{\emph{Class. Quant. Grav.}
  {\bfseries 23} (2006) 2435}
  [\href{https://arxiv.org/abs/gr-qc/0512039}{{\ttfamily gr-qc/0512039}}].

\bibitem{Harry:2006fi}
G.M.~Harry, P.~Fritschel, D.A.~Shaddock, W.~Folkner and E.S.~Phinney,
  \emph{{Laser interferometry for the big bang observer}},
  \href{https://doi.org/10.1088/0264-9381/23/15/008}{\emph{Class. Quant. Grav.}
  {\bfseries 23} (2006) 4887}.

\bibitem{Li:2003tv}
F.-Y.~Li, M.-X.~Tang and D.-P.~Shi, \emph{{Electromagnetic response of a
  Gaussian beam to high frequency relic gravitational waves in quintessential
  inflationary models}},
  \href{https://doi.org/10.1103/PhysRevD.67.104008}{\emph{Phys. Rev. D}
  {\bfseries 67} (2003) 104008}
  [\href{https://arxiv.org/abs/gr-qc/0306092}{{\ttfamily gr-qc/0306092}}].

\bibitem{Armengaud:2014gea}
E.~Armengaud et~al., \emph{{Conceptual Design of the International Axion
  Observatory (IAXO)}},
  \href{https://doi.org/10.1088/1748-0221/9/05/T05002}{\emph{JINST} {\bfseries
  9} (2014) T05002} [\href{https://arxiv.org/abs/1401.3233}{{\ttfamily
  1401.3233}}].

\bibitem{IAXO:2019mpb}
{\scshape IAXO} collaboration, \emph{{Physics potential of the International
  Axion Observatory (IAXO)}},
  \href{https://doi.org/10.1088/1475-7516/2019/06/047}{\emph{JCAP} {\bfseries
  06} (2019) 047} [\href{https://arxiv.org/abs/1904.09155}{{\ttfamily
  1904.09155}}].

\bibitem{Gross:1968in}
D.J.~Gross and R.~Jackiw, \emph{{Low-Energy Theorem for Graviton Scattering}},
  \href{https://doi.org/10.1103/PhysRev.166.1287}{\emph{Phys. Rev.} {\bfseries
  166} (1968) 1287}.

\bibitem{Gleisberg:2003ue}
T.~Gleisberg, F.~Krauss, K.T.~Matchev, A.~Schalicke, S.~Schumann and G.~Soff,
  \emph{{Helicity formalism for spin-2 particles}},
  \href{https://doi.org/10.1088/1126-6708/2003/09/001}{\emph{JHEP} {\bfseries
  09} (2003) 001} [\href{https://arxiv.org/abs/hep-ph/0306182}{{\ttfamily
  hep-ph/0306182}}].

\bibitem{deAquino:2011ix}
P.~de~Aquino, K.~Hagiwara, Q.~Li and F.~Maltoni, \emph{{Simulating graviton
  production at hadron colliders}},
  \href{https://doi.org/10.1007/JHEP06(2011)132}{\emph{JHEP} {\bfseries 06}
  (2011) 132} [\href{https://arxiv.org/abs/1101.5499}{{\ttfamily 1101.5499}}].

\bibitem{vanDam:1970vg}
H.~van Dam and M.J.G.~Veltman, \emph{{Massive and massless Yang-Mills and
  gravitational fields}},
  \href{https://doi.org/10.1016/0550-3213(70)90416-5}{\emph{Nucl. Phys. B}
  {\bfseries 22} (1970) 397}.

\end{thebibliography}\endgroup

\end{document}